\documentclass[aps,superscriptaddress,twocolumn,a4paper,showpacs]{revtex4}
\usepackage{graphicx}
\usepackage[all]{xy}
\usepackage{amsmath}
\usepackage{amssymb}
\usepackage{tensor}
\usepackage[colorlinks=true, pdfstartview=FitV, linkcolor=blue, citecolor=red, urlcolor=black, breaklinks=true]{hyperref}
\newcommand{\be}{\begin{equation}}
\newcommand{\ee}{\end{equation}}
\newcommand{\ben}{\begin{eqnarray}}
\newcommand{\een}{\end{eqnarray}}
\newcommand{\bes}{\begin{subequations}}
\newcommand{\ees}{\end{subequations}}
\def\bal#1\eal{\begin{align}#1\end{align}}

\newcommand{\arcsinh}{{\rm arcsinh}}
\newcommand{\sech}{{\rm sech}}
\newcommand{\csch}{{\rm csch}}
\newcommand{\LL}{{\cal L}}

\newcommand{\LX}{{\cal L}_X}

\newcommand{\vphi}{{\varphi}}

\begin{document}

\title{Vortices in vacuumless systems}
\author{M.A. Marques}\email{mam.matheus@gmail.com}\affiliation{Departamento de F\'\i sica, Universidade Federal da Para\'\i ba, 58051-970 Jo\~ao Pessoa, PB, Brazil}
\date{\today}
\begin{abstract}
We investigate the presence of vortex solutions in potentials without vacuum state. The study is conducted considering Maxwell and Chern-Simons dynamics. Also, we use a first order formalism that helps us to find the solutions and their respective electromagnetic fields and energy densities. As a bonus, we get to calculate the energy without knowing the explicit solutions. Even though the solutions present a large ``tail'' which goes far away from the origin, the magnetic flux remains a well defined topological invariant.
\end{abstract}

\maketitle
\section{Introduction}
In high energy physics, topological structures appear in a diversity of contexts and have been vastly studied over the years \cite{sut,vilenkin}. In spatial dimensions lower than three, the most known ones are kinks and vortices, which are static solutions of the equations of motion.

The simplest structures are kinks, which appear in $(1,1)$ spacetime dimensions by the action of scalar fields \cite{vachaspati}. Kinks connect the minima of the potential and have a topological character that assures its stability. However, it was shown in Ref.~\cite{vac} that topological defects may arise in potentials without a vacuum state, whose minima are located at infinity. Regarding the kink in the vacuumless system, it is asymptotically divergent and has infinite amplitude. Nevertheless, it is stable and can be associated to a topological charge by using a special definition for the topological current \cite{vbazeia}. Over the years, many papers have studied vacuumless topological defects in a diversity of contexts in high energy physics \cite{vac3,vac4,vac5,vac6,vac7,vac8,vac9,vac10,fermion}.

Potentials with extrema at infinity, similar to the ones we are going to study here, although inverted, also appear in classical mechanics \cite{perelomov}. In this scenario, if the energy is small enough, the motion is bounded. As the energy gets higher, the boundary values become far form each other, until the limit where they are infinitely separated. This limit distinguishes bounded and unbounded motion, so for sufficiently high values of the energy, the motion becomes unbounded. A similar situation happens in the interaction of a body with the gravitational potential, $V\propto-1/r$, which vanishes only at $r\to\infty$, when one calculates the escape velocity: the zero energy of the system describes the limit between bounded and unbounded motion. In high energy physics, vacuumless potentials arise in the massless limit of supersymmetric QCD due to non perturbative effects \cite{vacsusy}. They also appear in the cosmological context, where their energy densities could act as a cosmological constant that decreases slower than the densities of matter and radiation \cite{vaccosm1,vaccosm2}.

By working in $(2,1)$ spacetime dimensions one can find vortices. The first relativistic model that supports these objetcs was studied in Refs.~\cite{novortex,vega}, with the action of a complex scalar field coupled to a gauge field under the symmetry $U(1)$ in Maxwell dynamics. These structures are electrically neutral and engender a quantized flux which is conserved and works as a topological invariant. Their equations of motion are of second order with couplings between the fields; thus, they are hard to be solved. To simplify the problem, the BPS formalism was developed in Ref.~\cite{b,ps} for this model, which allowed for the presence of first order equations and the energy without knowing the explicit form of the solutions.

Models with the gauge field governed by the Maxwell dynamics, however, are not the only ones which support vortices solutions. One can also investigate these structures with the dynamics of the gauge field governed by the Chern-Simons term \cite{cs,jac1,jac2}. In this case, the vortex present a quantized flux, which also is topological invariant, and a quantized electric charge. The first studies of vortices in Chern-Simons dynamics are Refs.~\cite{cs1,cs2,cs3}; for more on this, see Ref.~\cite{dunne}.

The importance of vortices in high energy physics and in other areas of physics can be found in Refs.~\cite{vilenkin,sut,fradkin}. For instance, they may appear during the cosmic evolution of our Universe \cite{vilenkin} and in models that includes the so-called hidden sector, which is of interest in dark matter \cite{p1,p2,p3,p4}, by enlarging the symmetry to $U(1)\times U(1)$; see Refs.~\cite{hidden1,hidden2,hidden3}. Following this direction of enlarged symmetries, they are also present in $U(1)\times SO(3)$ models, with the addition of extra degrees of freedom to the vortex via the inclusion of a triplet scalar field, and in $U(1)\times Z_2$ models, with the inclusion of a neutral scalar field that acts as a source to the internal structure of the vortex \cite{vortexint}. Other motivations come from the context of condensed matter, where they may emerge in superconductors and in magnetic materials as magnetic domains \cite{mag}. They may also appear in dipolar Bose-Einstein condensates, where the atoms interact as dipole-dipole, which leads to the presence of non standard vortex structures \cite{cond1,cond2,cond3}.

Topological structures may be studied with generalized models \cite{babichev1,babichev2}. Vortices, in particular, firstly appeared in non canonical models in Refs.~\cite{gen1,gen2}. Since then, several works arised with other motivations. In the context of inflation, for instance, a model with a modified kinetic term was introduced in Ref.~\cite{kinf}. In this scenario, these models present distinct features from the standard case: they may not need a potential to drive the inflation. Moreover, generalized models were used in Refs.~\cite{cosm1,cosm2} as a tentative to explain why the universe is accelerated at a late stage of its evolution.

Non canonical models considering defect structures were severely investigated over the years \cite{kd1,kd2,kd3,kd4,werec,kd5,kd6,kd7,kd8,kd9,kd10,kd11,kd12,kd13}. Among the many investigations, a first order formalism was developed for some classes of non canonical models in Refs.~\cite{gen1,gen2,bazeiacs,kd10,fo2}. However, only in Ref.~\cite{godvortex} it was completely developed for any generalized model. An interesting fact is that, compact structures, which were firstly presented in Ref.~\cite{rosenau}, are possible to appear as Maxwell and Chern-Simons vortices only if generalized models are considered; see Refs.~\cite{compvortexm,compcs}. Non canonical models also allow for the presence of vortices that share the same field configuration and energy density, known as twinlike models \cite{vtwin}.

This work deals with a class of generalized Maxwell and Chern-Simons models that support vortex solutions in vacuumless systems. In Sec.~\ref{sec2}, we investigate the properties of vortices with Maxwell dynamics, including its first order formalism, and introduce two new models, one of them with analytical results. In Sec.~\ref{sec3} we conduct a similar investigation, however in the Chern-Simons scenario, also considering its first order formalism, and we introduce two new models. Finally, in Sec.~\ref{sec4} we present our ending comments and conclusions.

\section{Maxwell-Higgs Models}\label{sec2}
We deal with an action in $(2,1)$ flat spacetime dimensions for a complex scalar field and a gauge field governed by the Maxwell dynamics. We follow the lines of Ref.~\cite{godvortex} and write $S=\int d^3x \LL$, with the Lagrangian density given by
\be\label{lmax}
\LL = -\frac14 F_{\mu\nu}F^{\mu\nu} + K(|\vphi|)\overline{D_\mu \vphi}D^\mu\vphi -V(|\vphi|).
\ee
In the above equation, $\vphi$ denotes the complex scalar field, $A^\mu$ is the gauge field, $F_{\mu\nu}=\partial_\mu A_\nu-\partial_\nu A_\mu$ represents the electromagnetic strength tensor, $D_{\mu}=\partial _{\mu }+ieA_{\mu }$ stands for the covariant derivative, $e$ is the electric charge and $V(|\vphi|)$ is the potential, which is supposed to present symmetry breaking. The function $K(|\vphi|)$ is dimensionless and, in principle, arbitrary. Nevertheless, it has to admit solutions with finite energy. It is straightforward to show that $K(|\vphi|)=1$ gives the standard case considered in Ref.~\cite{novortex}. One may vary the action with respect to the fields $\vphi$ and $A_\mu$ to get the equations of motion
\bes\label{gmeom}
\bal
& D_\mu (K D^\mu\vphi)= \frac{\vphi}{2|\vphi|}\left(K_{|\vphi|}\overline{D_\mu \vphi}D^\mu\vphi -V_{|\vphi|} \right), \\ \label{meqs}
& \partial_\mu F^{\mu\nu}= J^\nu,
\eal
\ees
where the current is $J_\mu = ieK(|\vphi|)(\bar{\vphi}D_\mu \vphi-\vphi\overline{D_\mu\vphi})$ and $V_{|\vphi|} = dV/d|\vphi|$. Invariance under spacetime translations, $x^\mu\to x^\mu + a^\mu$, with $a^\mu$ constant, leads to the energy momentum tensor
\be\label{emtm}
T_{\mu\nu}= F_{\mu\lambda}\tensor{F}{^\lambda_\nu}+ K(|\vphi|)\left( \overline{D_\mu \vphi}D_\nu \vphi + \overline{D_\nu \vphi}D_\mu \vphi\right) - \eta_{\mu\nu} \LL.
\ee
In order to investigate vortex solutions in the model, we consider static configurations. As a consequence, the $\nu=0$ component of Eq.~\eqref{meqs} becomes an identity under the choice $A_0=0$. This makes the electric field vanish, so the vortex is electrically uncharged. Since we are dealing with two spatial dimensions, we define the magnetic field as $B=-F^{12}$. In this case, the surviving components of the energy momentum tensor \eqref{emtm} are
\bes
\begin{align}
T_{00} &= \frac{B^2}{2} + K(|\vphi|) \left|D_i\vphi\right|^2 + V(|\vphi|), \\
T_{12} &= K(|\vphi|) \left(  \overline{D_1 \vphi}D_2 \vphi + \overline{D_2 \vphi}D_1 \vphi\right), \\
T_{11} &= \frac{B^2}{2} + K(|\vphi|) \left(2\left|D_1\vphi\right|^2 -\left|D_i\vphi\right|^2\right) -V(|\vphi|), \\ 
T_{22} &= \frac{B^2}{2} + K(|\vphi|) \left(2\left|D_2\vphi\right|^2 -\left|D_i\vphi\right|^2\right) -V(|\vphi|).
\end{align}
\ees
The energy density is $\rho=T_{00}$ and the components $T_{ij}$ define the stress tensor. We then take the usual ansatz for vortex solutions
\bes\label{ansatz}
\begin{align}
\vphi(r,\theta) &=g(r)e^{i n\theta},\\
A_i &=-\epsilon_{ij} \frac{x^j}{er^2}[a(r)-n],
\end{align}
\ees
where $r$ and $\theta$ are the polar coordinates and $n=\pm1,\pm2,\ldots$ is the vorticity. The functions $g(r)$ and $a(r)$ must obey the boundary conditions
\be\label{bcond}
\begin{aligned}
g(0) &= 0, & a(0) &= n,\\
\lim_{r\to\infty} g(r) &= v, & \lim_{r\to\infty} a(r) &= 0.
\end{aligned}
\ee
In the above equations, $v$ is a parameter that is involved in the symmetry breaking of the potential. Considering the ansatz \eqref{ansatz}, the magnetic field becomes
\be\label{bm}
B(r) = -\frac{a^\prime}{er}.
\ee
By integrating it all over the space one can show that the flux is given by
\be\label{fluxm}
\begin{split}
\Phi &= \int d^2x B\\
     &= \frac{2\pi n}{e}.
\end{split}
\ee
Therefore, the magnetic flux is conserved and quantized by the vorticity $n$. As one knows, it is possible to introduce the conserved topological current
\be\label{topcurr}
j_T^\mu = \epsilon^{\mu\nu\lambda}\partial_\lambda A_\nu,
\ee
in which the component $j_T^0= B$ plays the role of a topological charge density. By integrating this, one can see that the flux \eqref{fluxm} plays an important role in the theory since it gives the topological charge of the system.

The equations of motion \eqref{gmeom} with the ansatz \eqref{ansatz} become
\bes\label{secansatzm}
\begin{align}\label{secansatzmg}
\frac{1}{r} \left(rK g^\prime\right)^\prime &= \frac{K a^2g}{r^2} + \frac12 V_{g}, \\\label{secansatzma}
r\left(\frac{a^\prime}{r} \right)^\prime &= 2e^2K ag^2.
\end{align}
\ees
Moreover, the components of the energy momentum tensor with the ansatz take the form
\bes
\begin{align}\label{rhom}
T_{00} &= \frac{{a^\prime}^2}{2e^2r^2} + K(g) \left({g^\prime}^2+\frac{a^2g^2}{r^2}\right) + V(g), \\
T_{12} &= K(g) \left( {g^\prime}^2 - \frac{a^2g^2}{r^2} \right) \sin(2\theta), \\ 
T_{11} &=  \frac{{a^\prime}^2}{2e^2r^2} + K(g) \bigg({g^\prime}^2(2\cos^2\theta-1) \nonumber\\
       &\hspace{4mm}+\frac{a^2g^2}{r^2}(2\sin^2\theta-1) \bigg) - V(g), \\ 
T_{22} &=   \frac{{a^\prime}^2}{2e^2r^2} + K(g) \bigg({g^\prime}^2(2\sin^2\theta-1) \nonumber\\
       &\hspace{4mm}+\frac{a^2g^2}{r^2}(2\cos^2\theta-1) \bigg) - V(g).
\end{align}
\ees

As was shown in Ref.~\cite{godvortex}, the stability against contractions and dilatations in the solutions requires the stressless condition. By setting $T_{ij}=0$, we get the first order equations
\be\label{fom}
g^\prime=\pm\frac{ag}{r} \quad\text{and}\quad -\frac{a^\prime}{er} = \pm\sqrt{2V(g)}.
\ee
The pair of equations for the upper signs are related to the lower signs ones by the change $a(r)\to-a(r)$. These equations are compatible with the equations of motion \eqref{secansatzm} if the potential and the function $K(|\vphi|)$ are constrained by
\be\label{constm}
\frac{d}{dg}\sqrt{2V(g)} = -2egK(g).
\ee
For $K(g)=1$, we have $V(|\vphi|) = e^2(v^2-|\vphi|^2)^2/2$, which is the standard case firstly studied in Ref.~\cite{novortex}. This constraint shows that generalized models are required to study different potentials and their correspondent vortexlike solutions from the ones of the standard case. The first-order equations \eqref{fom} also gives rise to the possibility of introducing an auxiliary function $W(a,g)$ in the form
\be\label{wm}
W(a,g) = -\frac{a}{e} \sqrt{2V(g)},
\ee
so the energy density is written as
\be
\rho = \frac{1}{r} \frac{dW}{dr}.
\ee
By integrating it all over the space, we get the energy
\be\label{ewm}
\begin{split}
	E &= 2\pi\left|W(a(\infty),g(\infty)) - W(a(0),g(0)) \right| \\
	  &= 2\pi\left|W(0,v) - W(n,0) \right|.
\end{split}
\ee
Thus, the energy of the stressless solutions may be calculated without knowing their explicit form. Below, by properly choosing $K(|\vphi|)$ and $V(|\vphi|)$ that satisfy the constraint in Eq.~\eqref{constm}, we show new models that engender a set of minima of the potential at infinity. Thus, we have $v=\infty$ in Eqs.~\eqref{bcond}. In order to prepare the model for numerical investigation, we work with dimensionless fields and consider unit vorticity, $n=1$, which requires the upper signs in the first order equations \eqref{fom}.

\subsection{First Model}
\begin{figure}[htb!]
\centering
\includegraphics[width=4.2cm]{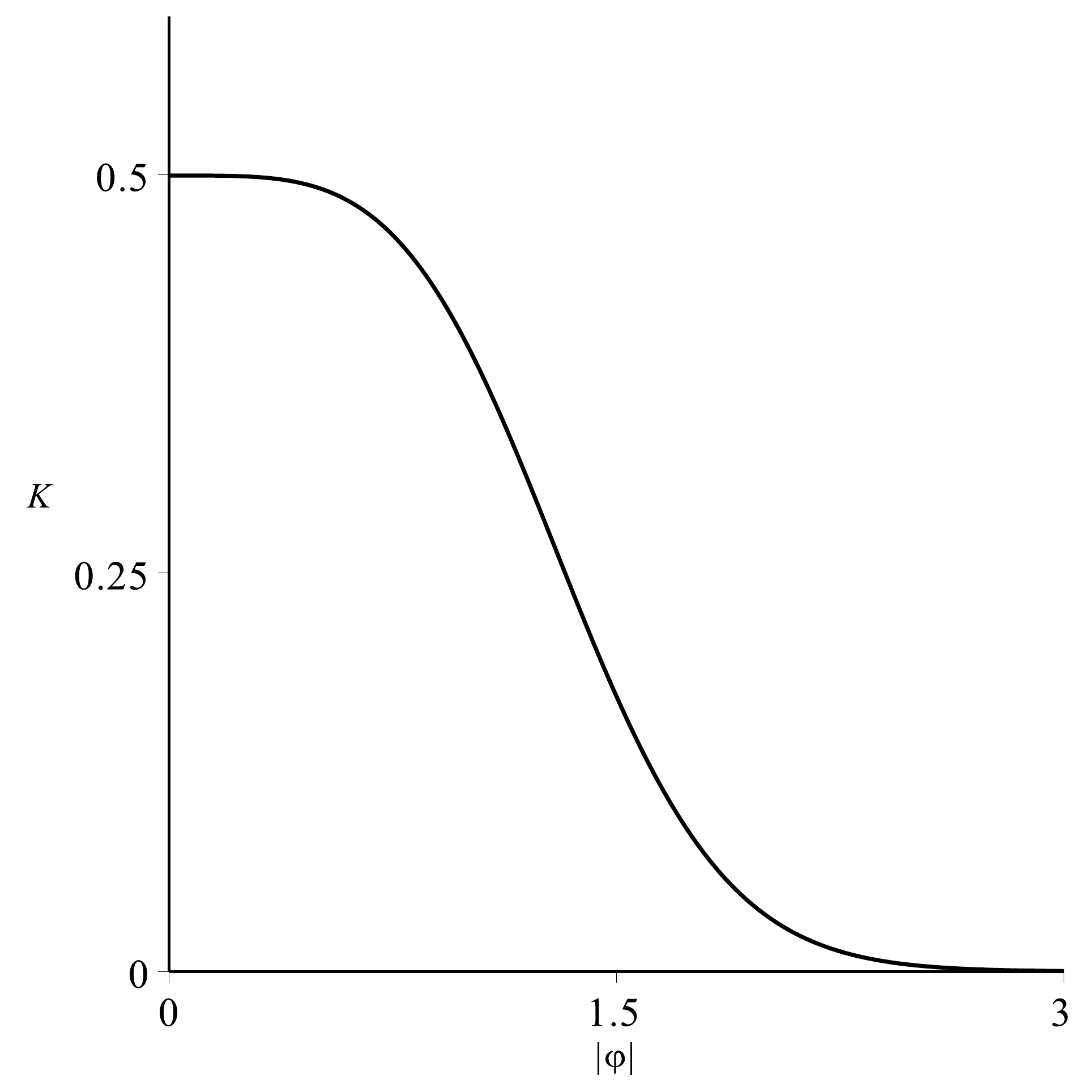}
\includegraphics[width=4.2cm]{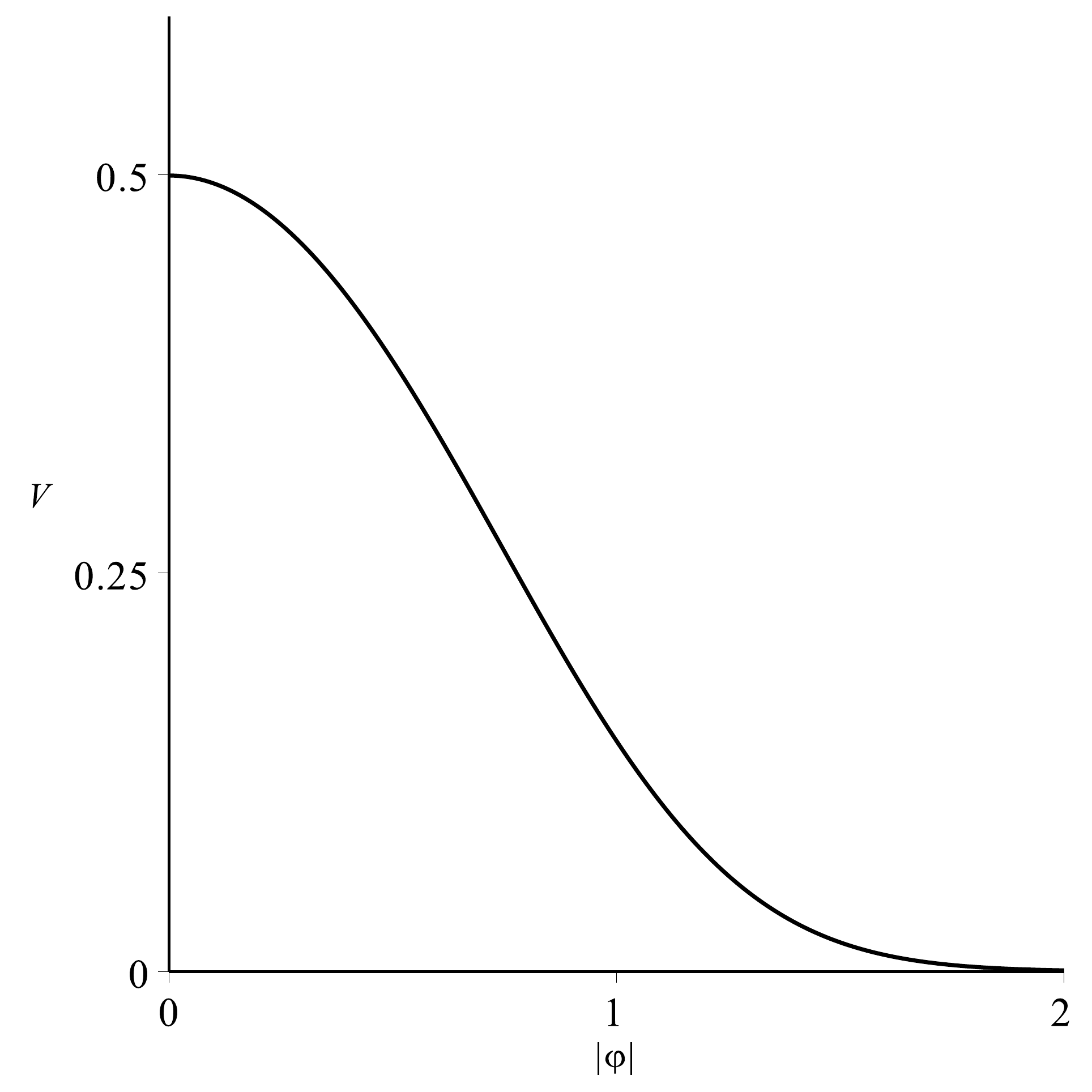}
\caption{The function $K(|\vphi|)$ (left) and the potential $V(|\vphi|)$ (right) given by Eqs.~\eqref{vk1m}.}
\label{fig1}
\end{figure}

The first example is given by the pair of functions
\bes\label{vk1m}
\bal
K(|\vphi|) &= \frac12\,\sech^2\left(\frac12\,|\vphi|^2\right), \\ 
V(|\vphi|) &= \frac12 \left(1-\tanh\left(\frac12\,|\vphi|^2\right)\right)^2.
\eal
\ees
The above potential does not present a vacuum state; that is the reason we call it vacuumless potential. However, since $V(\infty)=0$, we see the set of mimima of the potencial is located at infinity, which allows it to support vortex solutions. Its maximum is at $|\vphi_m|=0$, with $V(|\vphi_m|)=1/2$. In Fig.~\ref{fig1} we plot the above functions. We see that $K(|\vphi|)$, which is the function that controls the kinetic term of the model, behaves similarly to the potential $V(|\vphi|)$, having a maximum in the origin and its set of minima at infinity.

For this model, the first order equations \eqref{fom} become
\be\label{foex1m}
g^\prime=\frac{ag}{r} \quad\text{and}\quad \frac{a^\prime}{r} = -\left(1-\tanh\left(\frac{g^2}{2}\right)\right).
\ee
Near the origin, we can study the behavior of the solutions by taking $a(r)=1-a_0(r)$ and $g(r)=g_0(r)$ and going up to first order in $a_0(r)$ and $g_0(r)$. By substituting them in the above equations, we get that
\be\label{oriex1m}
a_0(r)\propto r^2 \quad\text{and}\quad g_0(r)\propto r.
\ee
It is worth commenting that, in this case, since the set of minima of the potential are at infinity, we see from the boundary conditions \eqref{bcond} that $g(r)$ is asymptotically divergent and has infinite amplitude, i.e., $g(r\to\infty)\to\infty$. Nevertheless, even though $g(r)$ goes to infinity, $a(r)$ still vanishes at infinity, similarly to what happens in the standard case.

\begin{figure}[htb!]
\centering
\includegraphics[width=4.2cm]{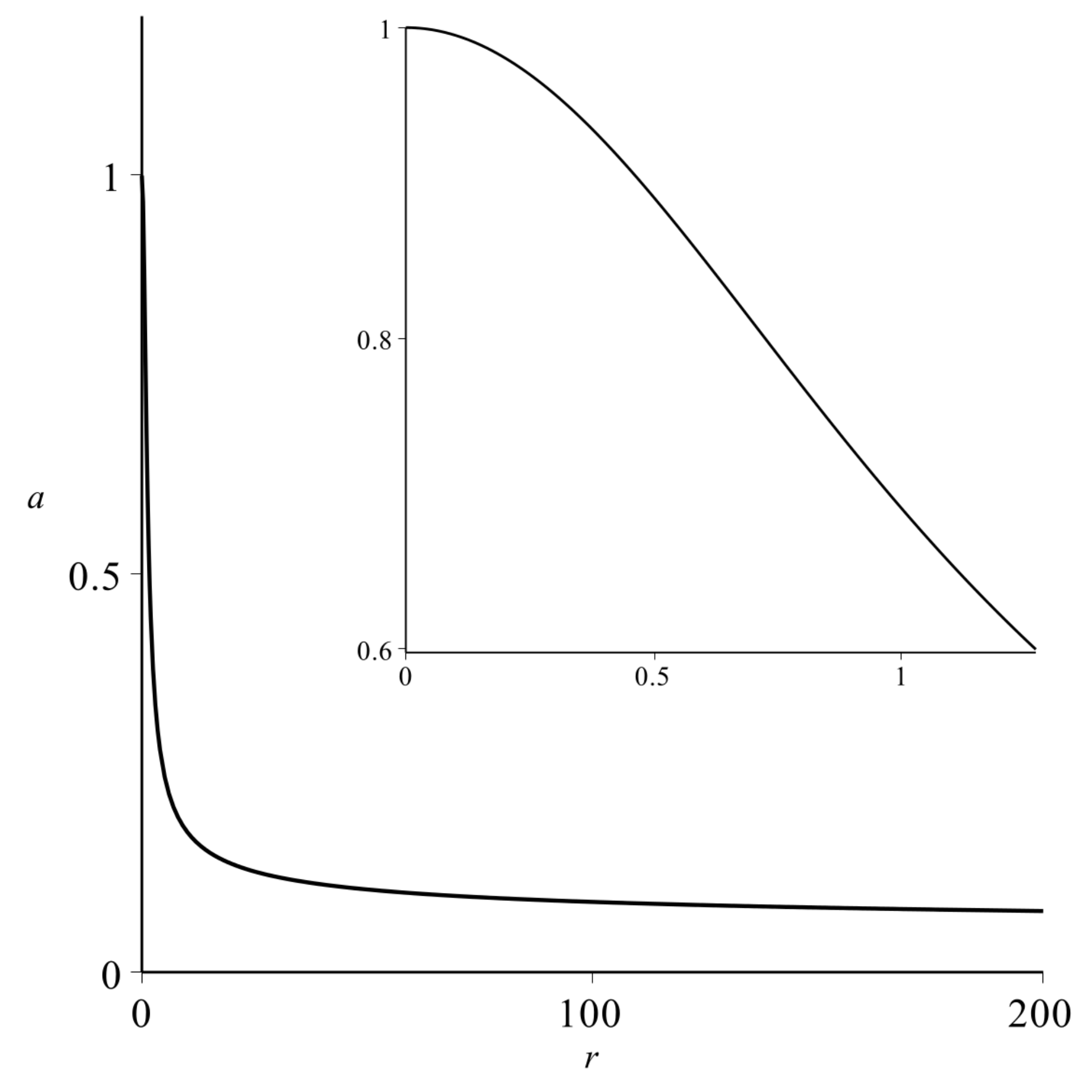}
\includegraphics[width=4.2cm]{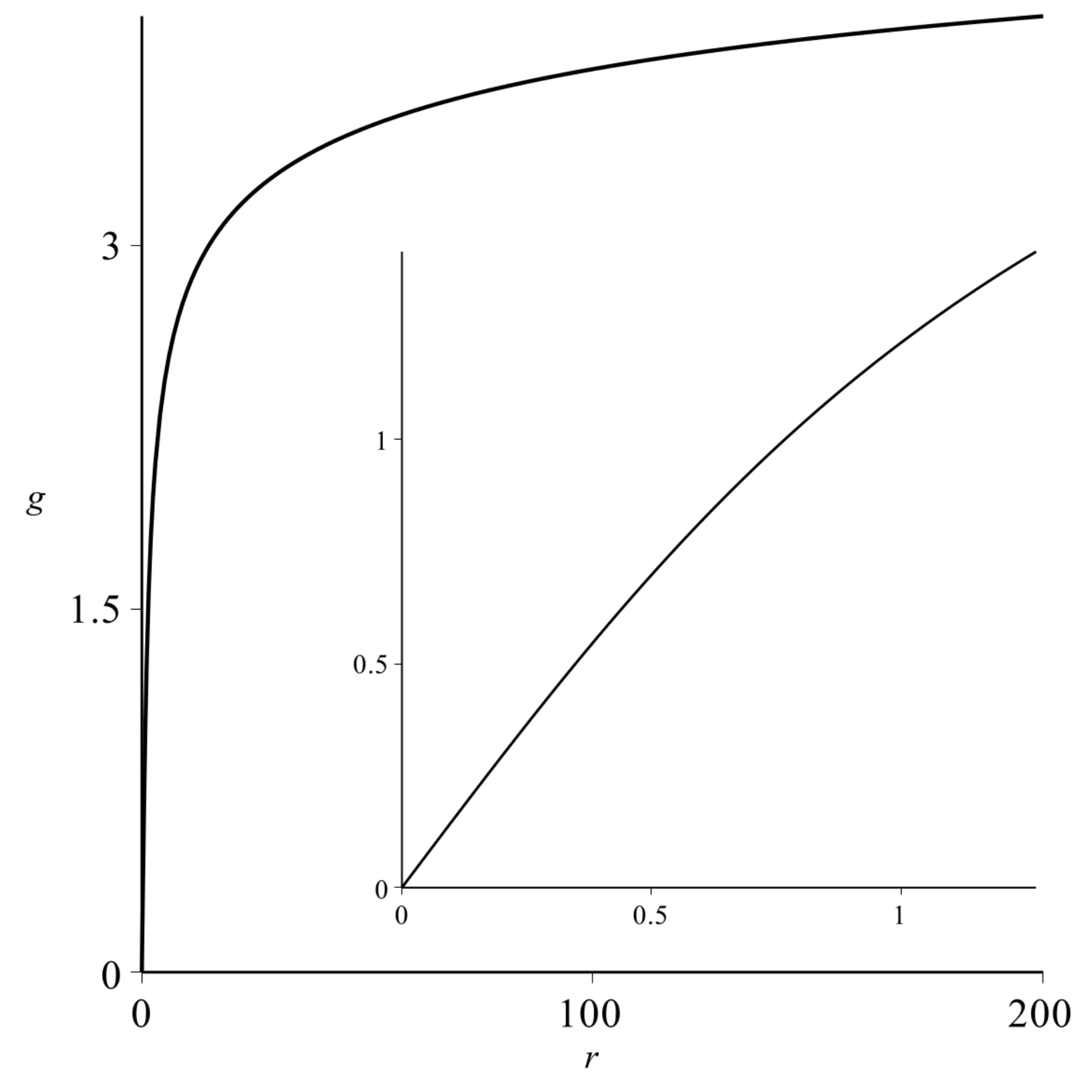}
\caption{The solutions $a(r)$ (left) and $g(r)$ (right) of Eqs.~\eqref{foex1m}. The insets show the behavior of the functions near the origin, in the interval $g\in[0,1.27]$.}
\label{fig2}
\end{figure}

Albeit Eqs.~\eqref{foex1m} are of first order, their nonlinearities makes the job of finding analytical solutions being very hard. Unfortunately, we have not been able to find them for these equations. Therefore, we must solve them by using numerical methods. In Fig.~\ref{fig2}, we plot the solutions of the above equations. Near the origin, we see that the functions vary as expected from Eq.~\eqref{oriex1m}. As $r$ increases, they tend to their boundary values very slow, which makes the tail of the solutions be present far away from the origin. This behavior is exactly the opposite from the one that appears in models which support compact vortices, in which the solutions attains their boundary values at a finite $r$ \cite{compvortexm}.

\begin{figure}[htb!]
\centering
\includegraphics[width=4.2cm]{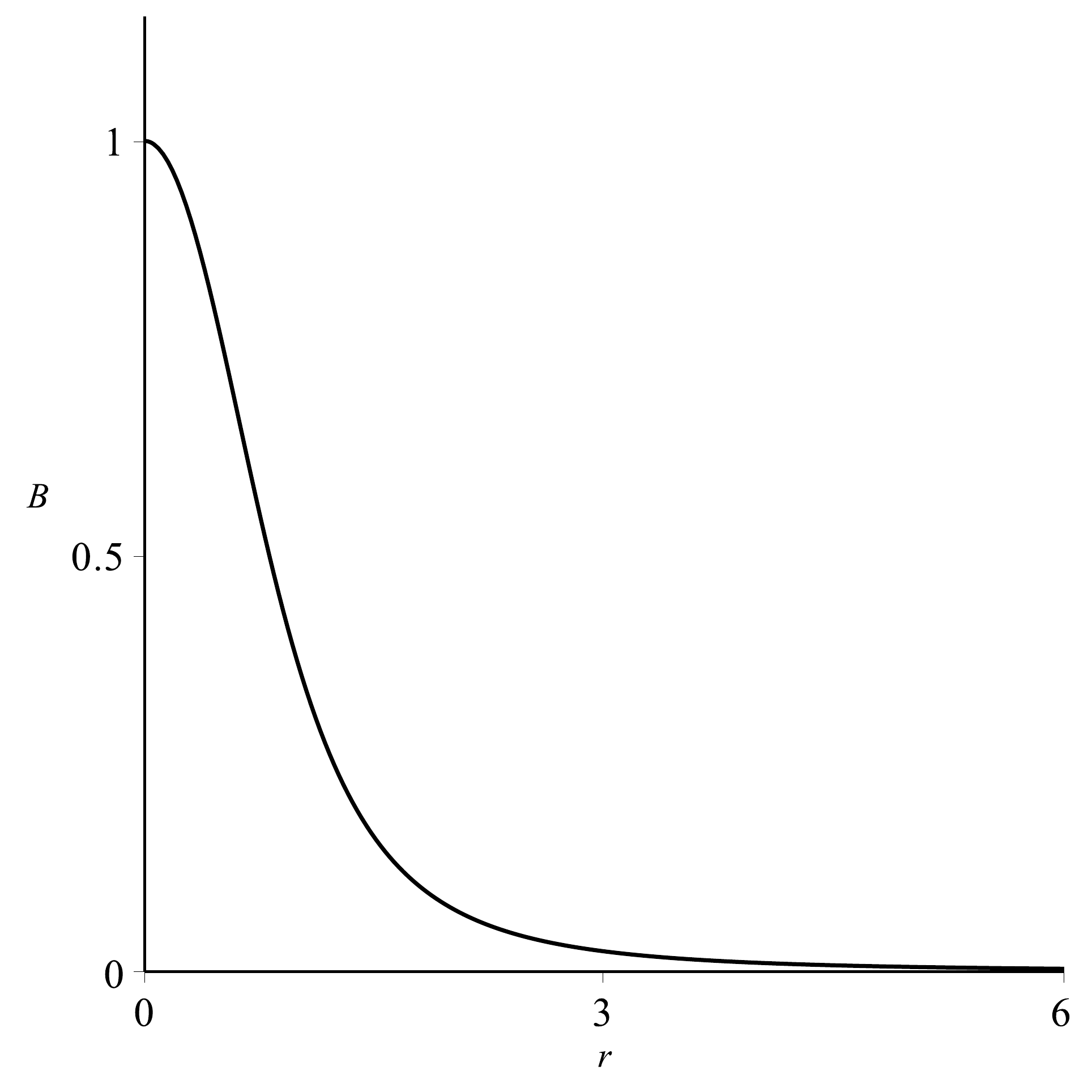}
\includegraphics[width=4.2cm]{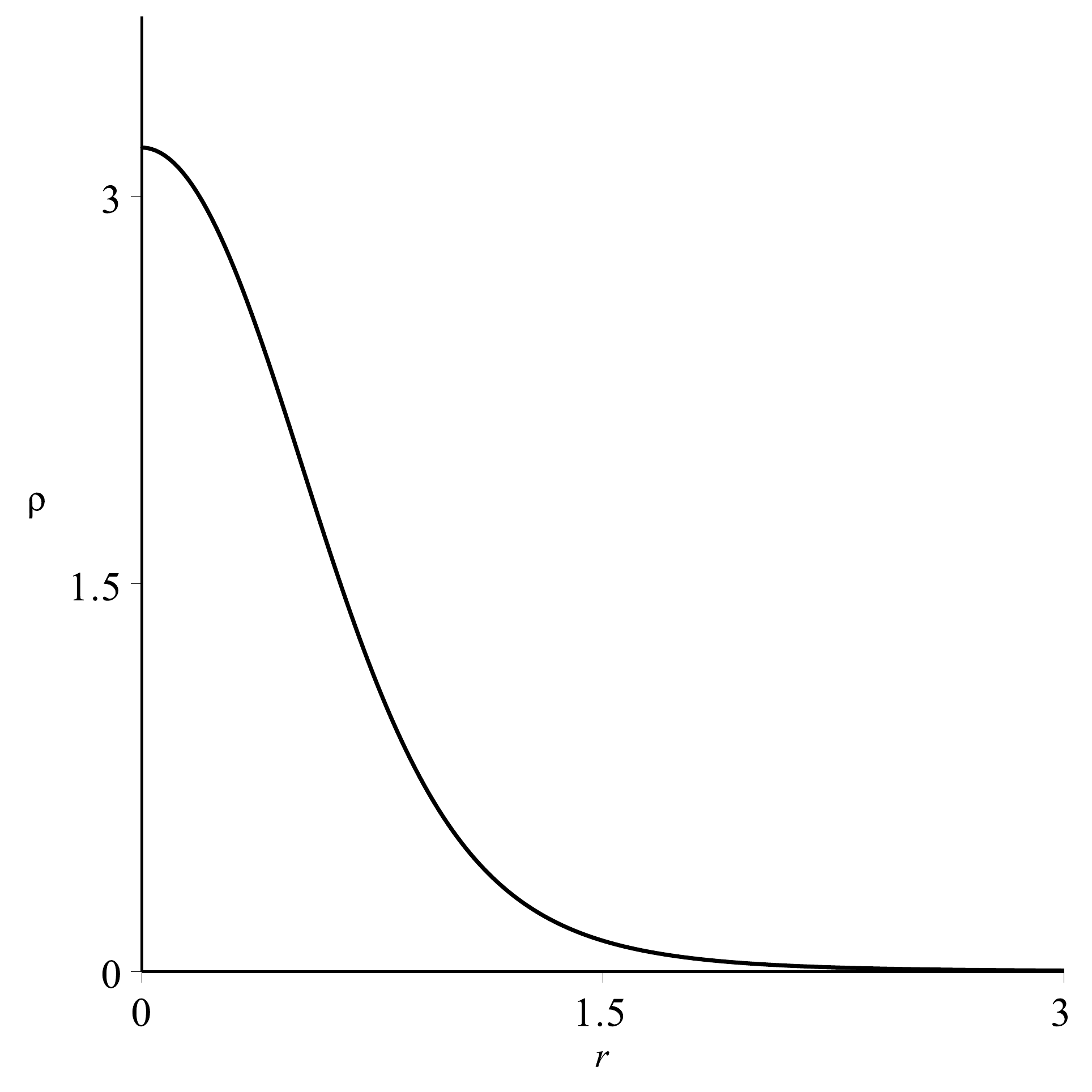}
\caption{The magnetic field (left) and the energy density (right) for the solutions of Eqs.~\eqref{foex1m}.}
\label{fig3}
\end{figure}

Before going further, we calculate the function $W(a,g)$, given by Eq.~\eqref{wm}:
\be\label{wex1m}
W(a,g) = -a+a\tanh\left(\frac{g^2}{2}\right).
\ee
By using Eq.~\eqref{ewm}, it is straightforward to show that the solutions of Eq.~\eqref{foex1m} have energy $E=2\pi$. The magnetic field is given by Eq.~\eqref{bm} and the energy density can be calculated from Eq.~\eqref{rhom}, which becomes
\be
\begin{aligned}
	\rho(r) = \;& \frac{{a^\prime}^2}{2e^2r^2} + \frac12\,\sech^2\left(\frac{g^2}{2}\right)  \left({g^\prime}^2+\frac{a^2g^2}{r^2}\right)\\
	          & +\frac12 \left(1-\tanh\left(\frac{g^2}{2}\right)\right)^2.
\end{aligned}
\ee
We then use our numerical solutions and plot the magnetic field and the energy density in Fig.~\ref{fig3}. One can see the large tail that the solutions have far away from the origin is less evident in the magnetic field and in the energy density. By numerical integration, one can show that the magnetic flux is well defined and given by $\Phi\approx2\pi$, as expected from Eq.~\eqref{fluxm}. Since the flux gives the topological charge associated to the vortex, this well defined behavior is different from the one for kinks in vacuumless systems, which require a special definition of topological current to get a topological character well defined \cite{vbazeia}. The numerical integration of the energy density all over the space gives energy $E\approx 2\pi$, which matches the value obtained with the using of the function $W(a,g)$ in Eq.~\eqref{wex1m}.

\subsection{Second Model}
\begin{figure}[htb!]
\centering
\includegraphics[width=4.2cm]{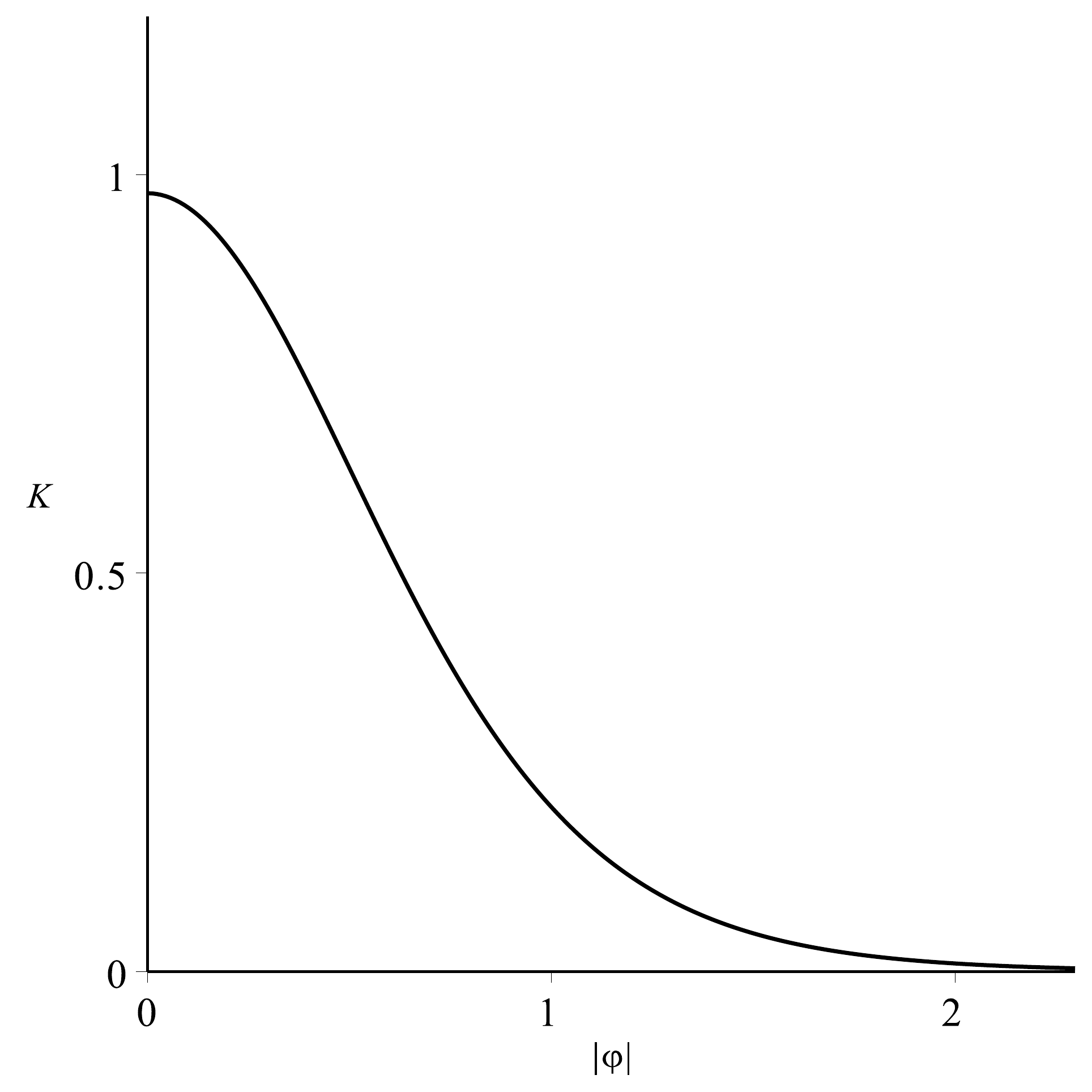}
\includegraphics[width=4.2cm]{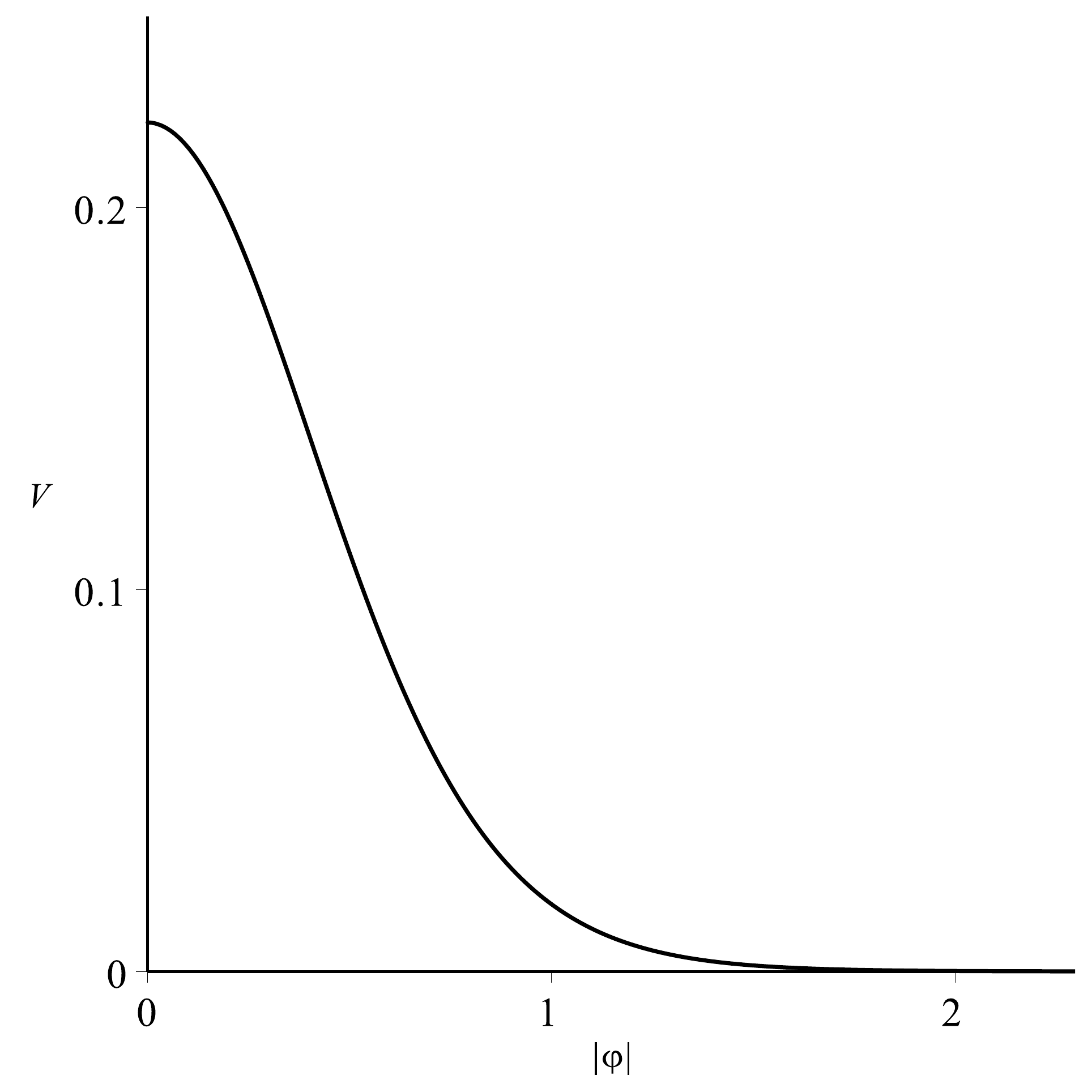}
\caption{The function $K(|\vphi|)$ (left) and the potential $V(|\vphi|)$ (right) given by Eqs.~\eqref{vk2m}.}
\label{fig4}
\end{figure}

Our second model arises from the pair of functions
\bes\label{vk2m}
\bal
K(|\vphi|) &=\frac{\left(2-(4-3S^2)\,C^2\right)S^2+ \left(2-3S^2\right)S\,C}{2|\vphi|^4},\\
V(|\vphi|) &= \frac{\left(1-S\,C\right)^2S^4}{2|\vphi|^4},
\eal
\ees
in which we have used the notation $S=\sech(|\vphi|)$ and $C=|\vphi|\,\csch(|\vphi|)$. Given the above expressions, one may wonder if these functions are finite in the origin. It is worth to investigate their behavior for $|\vphi|\approx0$, which is given by
\bes
\bal
K_{|\vphi|\approx0}(|\vphi|) &= \frac{44}{45} - \frac{1676}{945} |\vphi|^2 + \mathcal{O}\left(|\vphi^4| \right), \\
V_{|\vphi|\approx0}(|\vphi|) &= \frac29 - \frac{88}{135} |\vphi|^2 + \mathcal{O}\left(|\vphi^4| \right).
\eal
\ees
Then, they are regular at $|\vphi_m|=0$, which is a point of maximum with $V(|\vphi_m|)=2/9$. As in the previous model, this potential also is vacuumless. In Fig.~\ref{fig4}, we plot these functions. Notice that $K(|\vphi|)$ behaves similarly to the potential.

\begin{figure}[htb!]
\centering
\includegraphics[width=4.2cm]{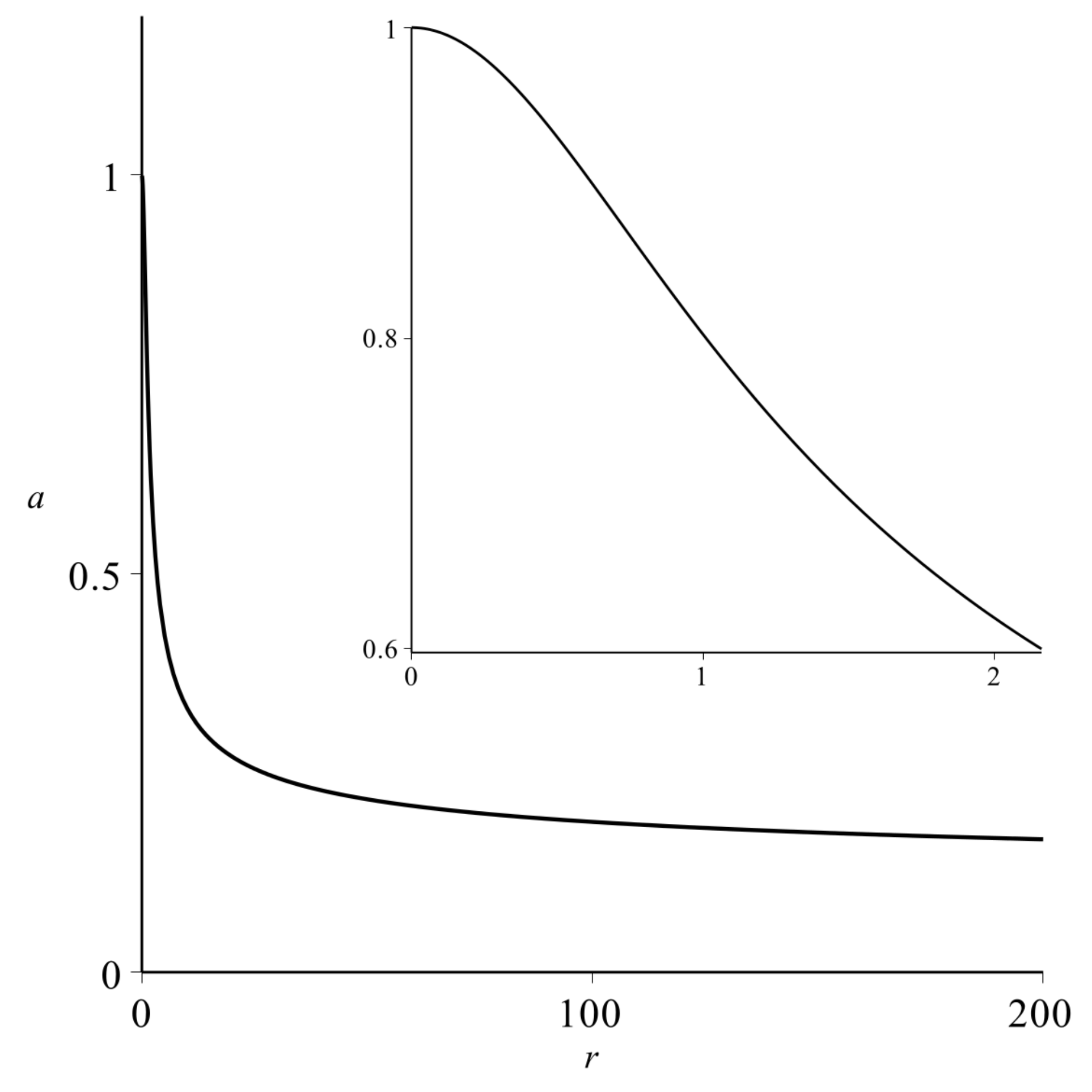}
\includegraphics[width=4.2cm]{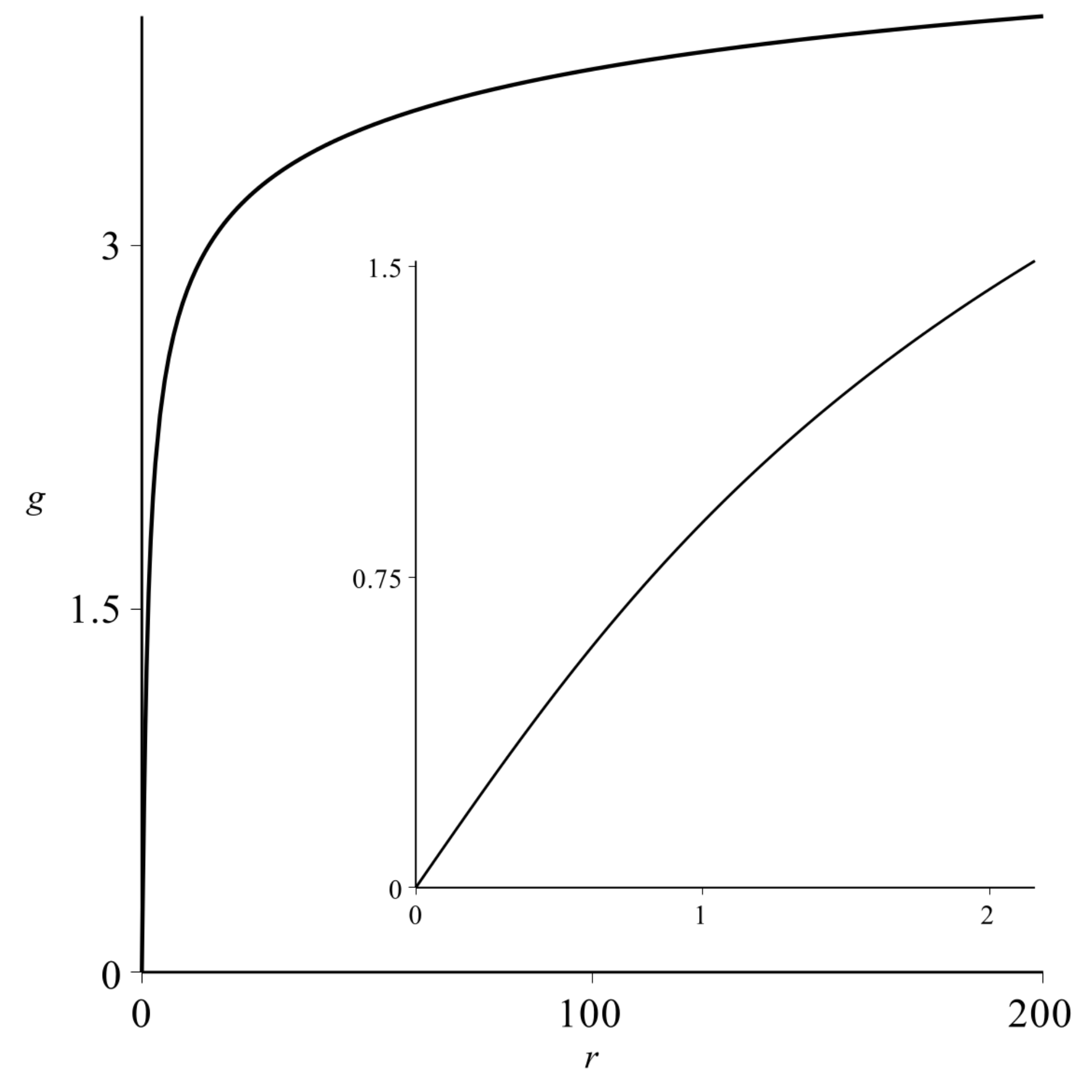}
\caption{The solutions $a(r)$ (left) and $g(r)$ (right) as in Eqs.~\eqref{solex2m}. The insets show the behavior of the functions near the origin, in the interval $g\in[0,2.16]$.}
\label{fig5}
\end{figure}

In this case, the first order equations \eqref{fom} take the form
\bes\bal
g^\prime &=\frac{ag}{r},\\
\frac{a^\prime}{r} &= - \frac{\left(1-g\,\sech(g)\,\csch(g)\right)\sech^2(g)}{g^2}.
\eal
\ees
The behavior near the origin can be studied by considering $a(r)=1-a_0(r)$ and $g(r)=g_0(r)$ and going up to first order in $a_0(r)$ and $g_0(r)$. Plugging them in the above equations, we get the same behavior of Eq.~\eqref{oriex1m}. The above equations admit the solutions
\bes\label{solex2m}
\bal
g(r) &= \arcsinh(r),\\
a(r) &= \frac{r}{\sqrt{1+r^2}\,\arcsinh(r)}.
\eal
\ees
Therefore, as expected, $g(r)$ goes to infinity and $a(r)$ vanishes very slowly as $r$ increases. Then, as in the previous model, the tail of the solutions is present even for large distances from the origin. This behavior is shown in Fig.~\ref{fig5}, in which we plot these solutions.

\begin{figure}[htb!]
\centering
\includegraphics[width=4.2cm]{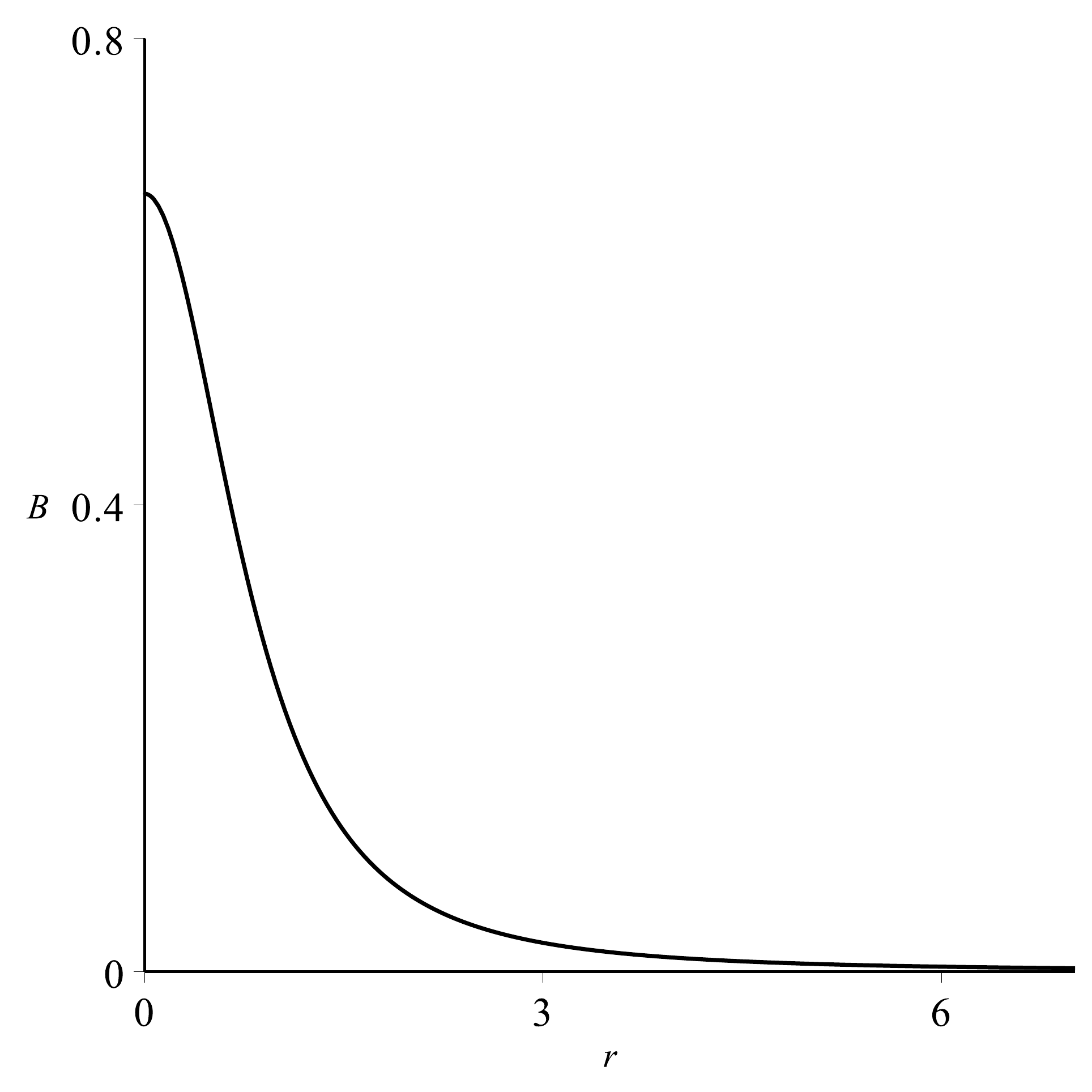}
\includegraphics[width=4.2cm]{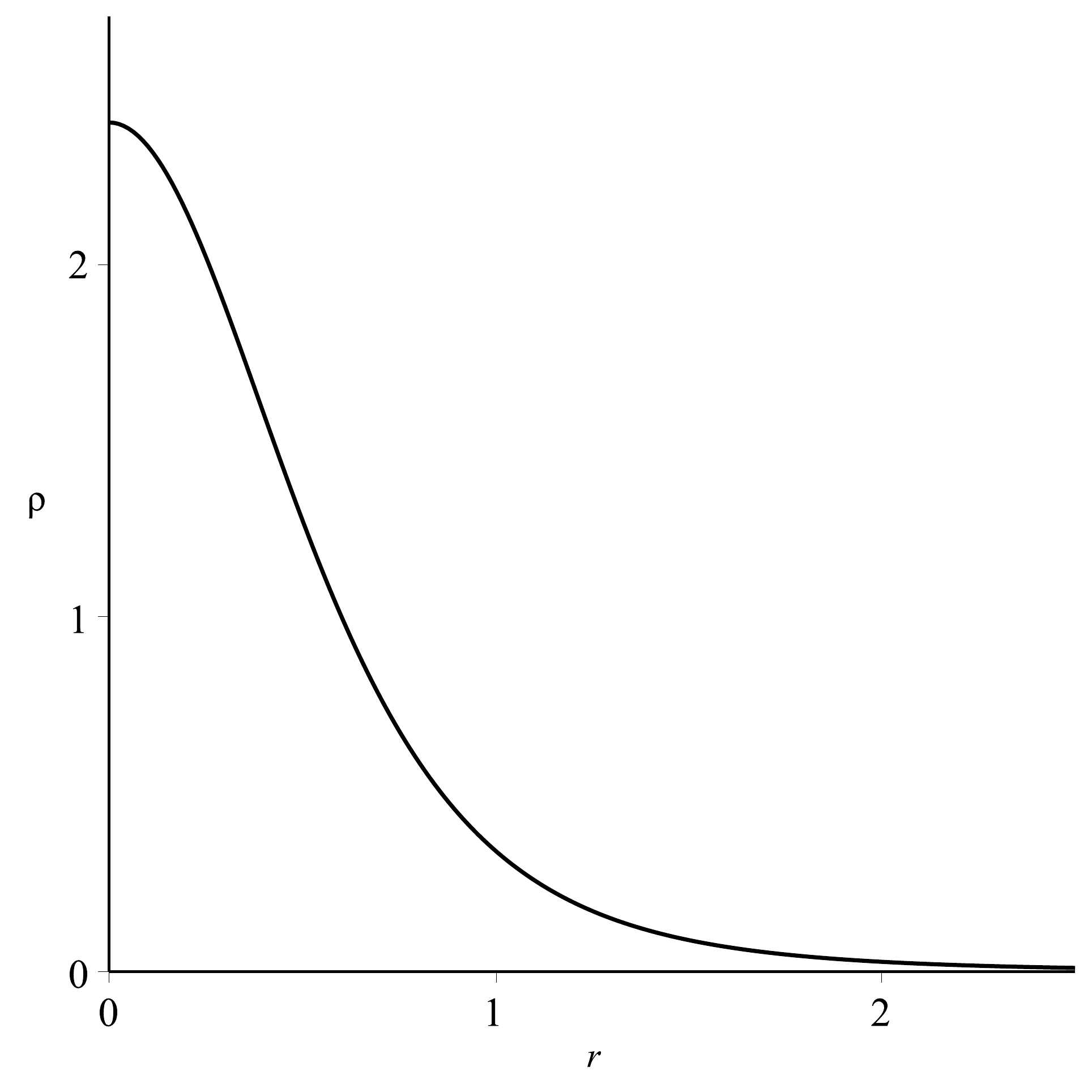}
\caption{The magnetic field in Eq.~\eqref{bex2m} (left) and the energy density as in Eq.~\eqref{rhoex2m} (right).}
\label{fig6}
\end{figure}

In this case, $W(a,g)$, given by Eq.~\eqref{wm}, takes the form
\be\label{wex2m}
W(a,g) = -\frac{a\left(1-g\,\sech(g)\,\csch(g)\right)\sech^2(g)}{g^2}.
\ee
Then, from Eq.~\eqref{ewm}, the solutions \eqref{solex2m} have energy $E=4\pi/3$. Since we have the analytical solutions in this case, we can calculate the magnetic field from Eq.~\eqref{bm} and the energy density from Eq.~\eqref{rhom} to get
\bes
\bal\label{bex2m}
B(r) &= \frac{r\,\sqrt{1+r^2} - \arcsinh(r)}{r\,\arcsinh^2(r) \left(1+r^2\right)^{3/2}},\\ \label{rhoex2m}
\rho(r) &= \frac{\left(2r^2-3\right)\sqrt{1+r^2}\,\arcsinh(r) - 4\,r\,\arcsinh^2(r)}{r\,\arcsinh^4(r)\left(1+r^2\right)^3} \nonumber\\
     &\hspace{4mm}+\frac{3\,r\left(r^2+1\right)}{r\,\arcsinh^4(r)\left(1+r^2\right)^3}.
\eal
\ees
In Fig.~\ref{fig6}, we plot the magnetic field and the energy density.
A direct integration of the magnetic field \eqref{bex2m} gives exactly the flux in Eq.~\eqref{fluxm}. The energy obtained by an integration of the energy density \eqref{rhoex2m} gives the same value obtained by the using of the auxiliary function $W(a,g)$ in Eq.~\eqref{wex2m}, that is, $E=4\pi/3$. As in the previous model, the long tail of the solutions does not seem to modify the flux of the vortex, which remains as in Eq.~\eqref{fluxm}. Then, the topological current \eqref{topcurr} is a definition that leads to a well behaved topological charge.

\section{Chern-Simons-Higgs Models}\label{sec3}
In order to investigate the presence of vortices with the Chern-Simons dynamics, we consider the action $S=\int d^3x \LL$ for a complex scalar field and a gauge field. Here, we study the class of generalized models presented in Ref.~\cite{bazeiacs}
\begin{equation}\label{lcs}
\LL = \frac{\kappa}{4}\epsilon^{\alpha\beta\gamma}A_\alpha F_{\beta\gamma} + K(|\vphi|)\overline{D_\mu \vphi}D^\mu\vphi -V(|\vphi|).
\end{equation}
In the above expression, $\vphi$, $A^\mu$, $e$, $D_{\mu}=\partial _{\mu }+ieA_{\mu}$, $F_{\mu\nu}=\partial_\mu A_\nu-\partial_\nu A_\mu$ and $V(|\vphi|)$ have the same meaning of the previous section. Here, $\kappa$ is a constant. Regarding the dimensionless function $K(|\vphi|)$, it is in principle arbritrary. The only restriction for it is to provide solutions with finite energy. The standard case is given by $K(|\vphi|)=1$ and was studied in Ref.~\cite{cs2}. Here, we consider $A^\mu = (A^0,\vec{A})$. Thus, the electric and magnetic fields are
\be\label{eb1cs}
E^i = F^{i0} = -\dot{A}^i - \partial_i A^0 \quad\text{and}\quad B = -F^{12},
\ee
with the dot meaning the temporal derivative and $(E_x,E_y)\equiv E^i$, where $i=1,2$. The equations of motion for the scalar and gauge fields read
\bes\label{eomcs}
\begin{align}
 D_\mu (K D^\mu\vphi)&= \frac{\vphi}{2|\vphi|}\left(K_{|\vphi|}\overline{D_\mu \vphi}D^\mu\vphi -V_{|\vphi|} \right), \\ \label{cseqscs}
 \frac{\kappa}{2} \epsilon^{\lambda\mu\nu}F_{\mu\nu} &= J^\lambda,
\end{align}
\ees
where the current is $J_\mu = ieK(|\vphi|)(\bar{\vphi}D_\mu \vphi-\vphi\overline{D_\mu\vphi})$. Since the Chern-Simons term in the Lagrangian density \eqref{lcs} is metric-free, it does not contribute to the energy momentum tensor, which has the form  
\bal\label{emtcs}
T_{\mu\nu}&=K(|\vphi|)\left( \overline{D_\mu \vphi}D_\nu \vphi + \overline{D_\nu \vphi}D_\mu \vphi\right) \nonumber\\
          &\hspace{4mm}- \eta_{\mu\nu} \left( K(|\vphi|)\overline{D_\lambda \vphi}D^\lambda\vphi -V(|\vphi|) \right).
\eal
We now consider static solutions and the same ansatz of Eqs.~\eqref{ansatz} with the boundary conditions \eqref{bcond}. This makes the electric and magnetic fields in Eq.~\eqref{eb1cs} have the form 
\be\label{eb2cs}
E^i= -\partial_i A_0 \quad\text{and}\quad B = -\frac{a^\prime}{er}.
\ee
The magnetic flux can by calculated and it is given by Eq.~\eqref{fluxm}, which shows that it is quantized and conserved. Therefore, the Maxwell and Chern-Simons vortices share the same magnetic flux. Furthermore, we can also consider the topological current as in Eq.~\eqref{topcurr} to show that the topological charge is given by the magnetic flux. We must be careful, though, with the temporal component of the gauge field, $A_0$. In this case, the Gauss' law that appears in Eq.~\eqref{cseqscs} for $\lambda=0$ is not solved for $A_0=0$. Moreover, $A_0$ is not an independent function; one can show that it is given by
\be\label{A0}
A_0 = \frac{\kappa}{2e^2} \frac{B}{|\vphi|^2K(|\vphi|)}.
\ee
Since the electric field does not vanish, Chern-Simons vortices engender electric charge, given by
\be
\begin{split}
	Q &= \int d^2x J^0 \\
	  &= -\kappa \Phi.
\end{split}
\ee
Therefore, given the quantized magnetic flux \eqref{fluxm}, the electric charge is also quantized by the vorticity $n$. The equations of motion \eqref{eomcs} with the ansatz \eqref{ansatz} and $A_0=A_0(r)$, are given by
\bes\label{eomansatzcs}
\begin{align}
\frac{1}{r} \left(rK g^\prime\right)^\prime + K g \left(e^2 A_0^2-\frac{a^2}{r^2} \right)+& \nonumber\\
+ \frac12 \left(\left(e^2g^2A_0^2-{g^\prime}^2-\frac{a^2g^2}{r^2}\right)K_{g} -V_{g} \right) &= 0, \\ \label{a0csansatz}
 \frac{a^\prime}{r} + \frac{2K e^3g^2 A_0}{\kappa} &= 0, \\
 {A_0^\prime} + \frac{2K ea g^2}{\kappa r} &= 0.
\end{align}
\ees
The components of the energy momentum tensor \eqref{emtcs} with the ansatz \eqref{ansatz} read
\bes\label{tmunucs}
\begin{align}\label{rhocs}
T_{00} & = \frac{\kappa^2}{4e^4} \frac{{a^\prime}^2}{r^2 g^2K(g)} + \left({g^\prime}^2+\frac{a^2g^2}{r^2}\right)K(g) + V(g), \\
T_{01} &= -\frac{2K(g) e ag^2 A_0 \sin{\theta}}{r},  \\
T_{02} &= \frac{2\LX e ag^2 A_0 \cos{\theta}}{r}, \\
T_{12} &= K(g) \left( {g^\prime}^2 - \frac{a^2g^2}{r^2} \right) \sin(2\theta), \\ 
T_{11} &= K(g) \bigg(e^2g^2A_0^2 +{g^\prime}^2(2\cos^2\theta-1) \nonumber\\
       &\hspace{4mm}+\frac{a^2g^2}{r^2}(2\sin^2\theta-1) \bigg) - V(g), \\ 
T_{22} &= K(g) \bigg(e^2g^2A_0^2 + {g^\prime}^2(2\sin^2\theta-1) \nonumber\\
       &\hspace{4mm}+\frac{a^2g^2}{r^2}(2\cos^2\theta-1) \bigg) - V(g).
\end{align}
\ees
The equations of motion \eqref{eomansatzcs} are coupled differential equations of second order. To simplify the problem and get first order equations, we follow Ref.~\cite{godvortex} and take the stressless condition, $T_{ij}=0$. This leads to
\be
g^\prime = \frac{ag}{r} \quad\text{and}\quad e^2A_0^2g^2K(g)=V(g).
\ee
We can combine this with Gauss' law \eqref{a0csansatz} to get the two first order equations
\be\label{focs}
g^\prime = \frac{ag}{r} \quad\text{and}\quad \frac{a^\prime}{r} =-\frac{2e^2g}{\kappa}\sqrt{KV},
\ee
in which the functions $K(|\vphi|)$ and $V(|\vphi|)$ are constrained by
\be\label{KVcs}
\frac{d}{dg}\left(\sqrt{\frac{V}{g^2K}}\right)=-\frac{2e^2}{\kappa}g K.
\ee
For $K(|\vphi|)=1$ we have the potential given by $V(|\vphi|) = e^4|\vphi|^2(1-|\vphi|^2)^2/\kappa^2$, which was studied in Ref.~\cite{cs2}. The first order equations allow us to introduce an auxiliary function $W(a,g)$, given by
\be\label{wcs}
W(a,g) = -\frac{\kappa\, a}{e^2 g} \sqrt{\frac{V(g)}{K(g)}},
\ee
and write the energy density in Eq.~\eqref{rhocs} as
\be
\rho = \frac{1}{r}\frac{dW}{dr}.
\ee
By integrating it, we get the energy 
\be\label{ewcs}
\begin{split}
	E &= 2\pi\left|W(a(\infty),g(\infty)) - W(a(0),g(0)) \right| \\
	  &= 2\pi\left|W(0,v) - W(n,0) \right|.
\end{split}
\ee
This formalism allows us to calculate the energy of the stressless solutions without knowing their explicit form. As done in the latter section, for simplicity, we neglect the parameters and work with unit vorticity, $n=1$. Next, we present models in the above class that admit vortices in potentials with minima located at infinity, i.e., $v\to\infty$ in the boundary conditions \eqref{bcond}.

\subsection{First Model}
\begin{figure}[htb!]
\centering
\includegraphics[width=4.2cm]{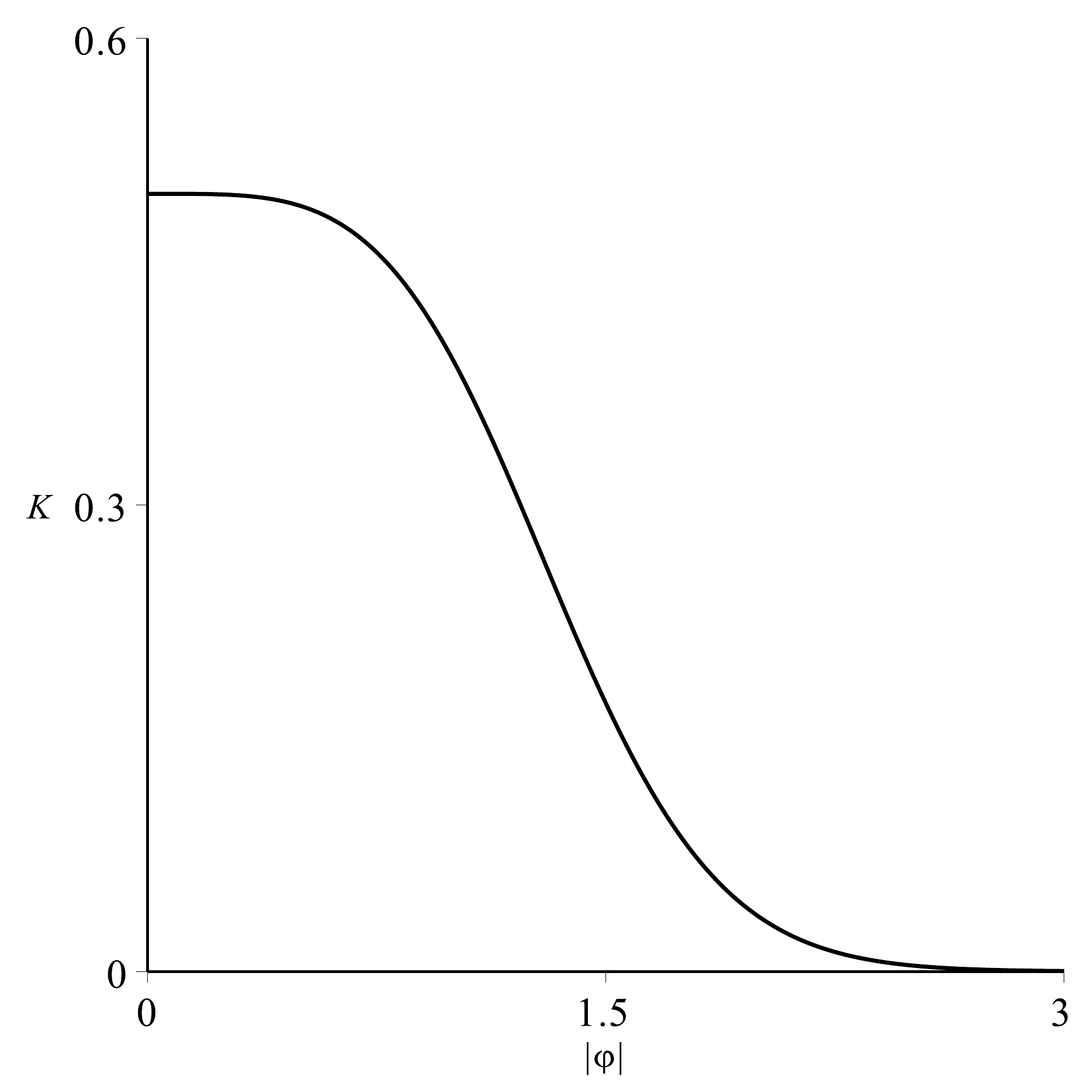}
\includegraphics[width=4.2cm]{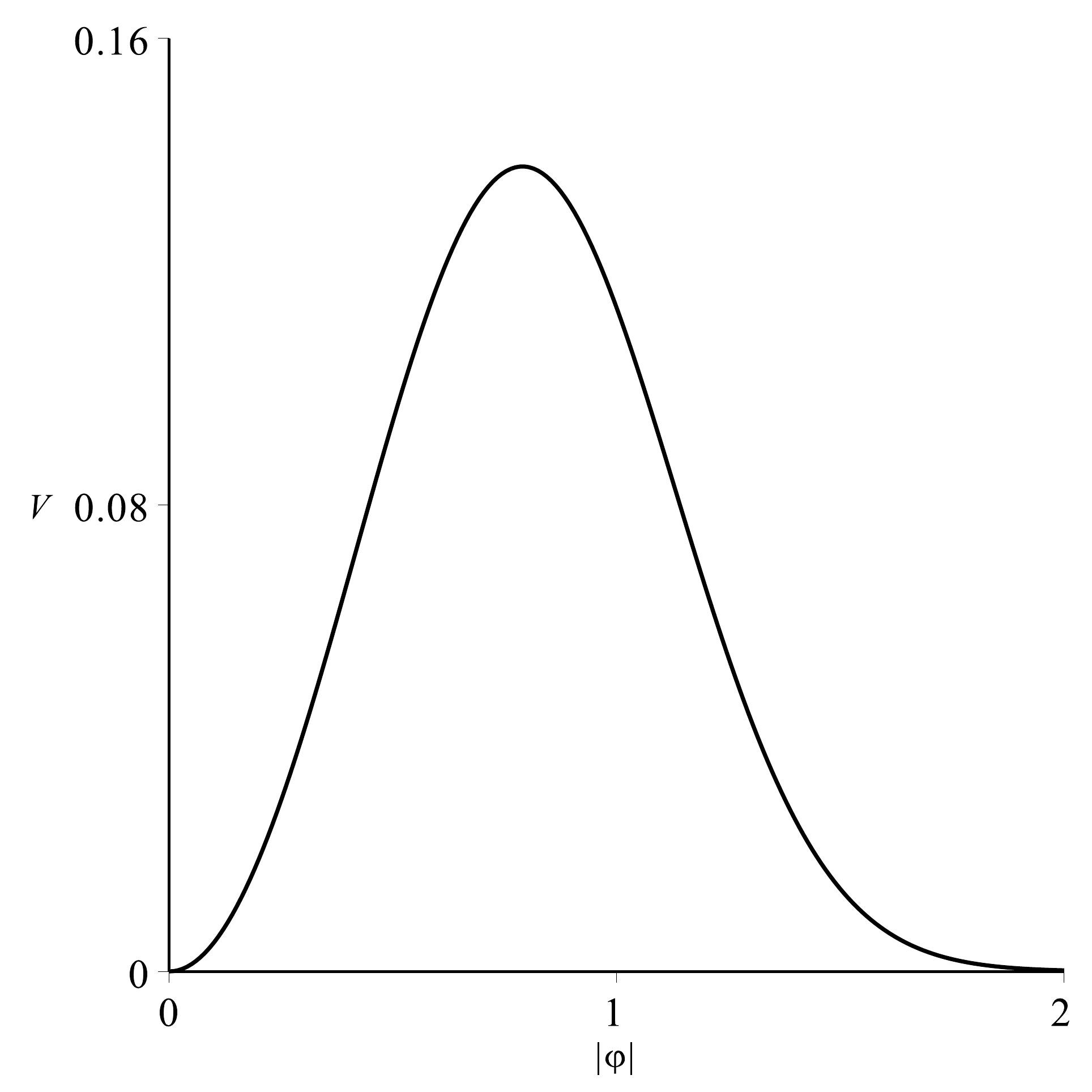}
\caption{The function $K(|\vphi|)$ (left) and the potential $V(|\vphi|)$ (right) given by Eqs.~\eqref{vk1cs}.}
\label{fig7}
\end{figure}

To start the investigation with the Chern-Simons dynamics, we consider the same $K(|\vphi|)$ of Eq.~\eqref{vk1m} but with other potential in order to satisfy the constraint \eqref{KVcs}. We then take
\bes\label{vk1cs}
\bal
K(|\vphi|) &= \frac12\,\sech^2\left(\frac12\,|\vphi|^2\right), \\ 
V(|\vphi|) &= \frac12|\vphi|^2\sech^2\left(\frac12|\vphi|^2 \right) \left(1-\tanh\left(\frac12\,|\vphi|^2\right)\right)^2.
\eal
\ees
These functions are plotted in Fig.~\ref{fig7}. The potential presents a minimum at $|\vphi|=0$ and a set of minima at $|\vphi|\to\infty$. Its maximum is located at $|\vphi_m|\approx 0.79$, such that $V(|\vphi_m|) \approx 0.14$. Furthermore, even though the function $K(|\vphi|)$ is the same of Eq.~\eqref{vk1m} in Maxwell dynamics, we see its corresponding potential has a completely different behavior near the origin in the Chern-Simons dynamics, with a minimum instead of a maximum at $|\vphi|=0$.

\begin{figure}[htb!]
\centering
\includegraphics[width=4.2cm]{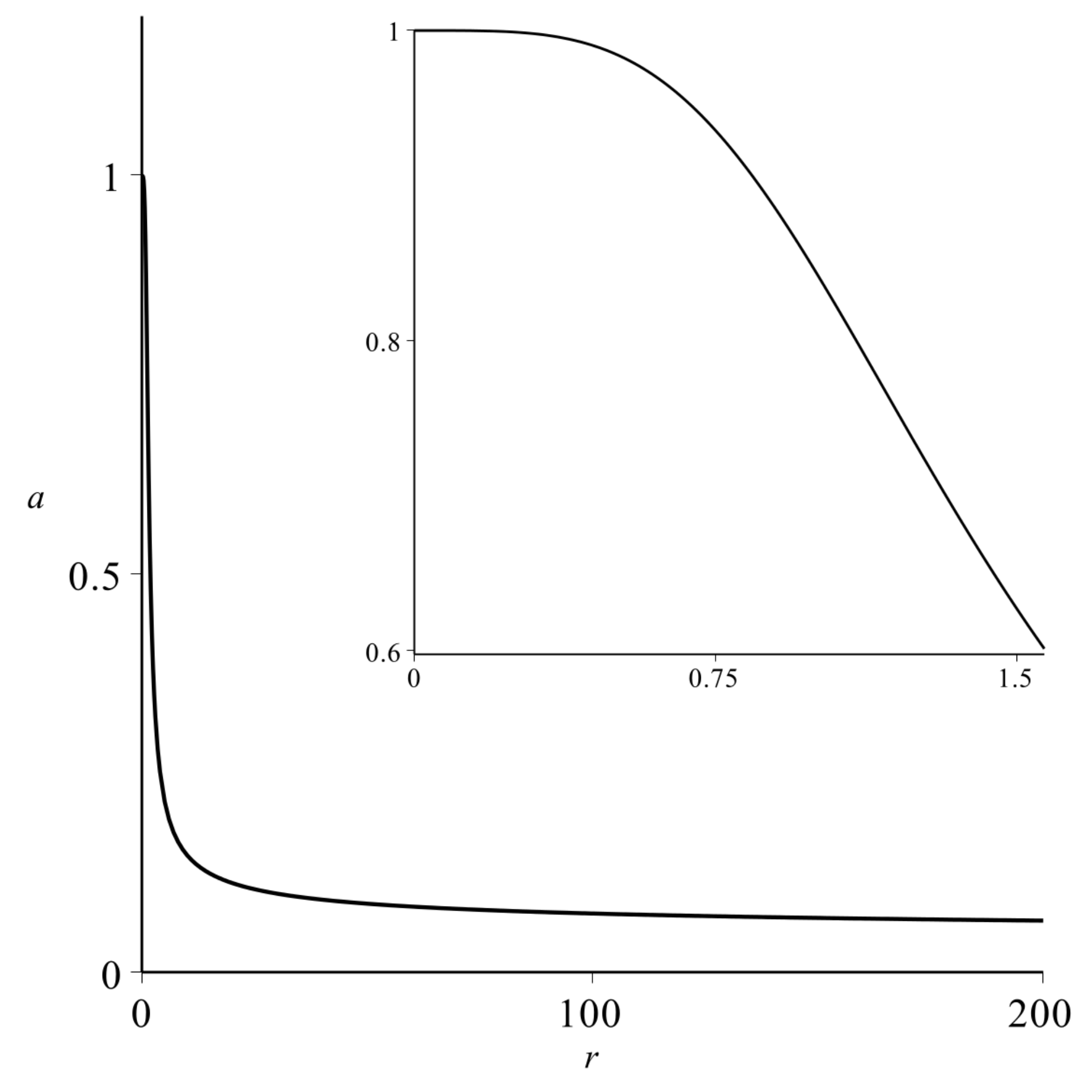}
\includegraphics[width=4.2cm]{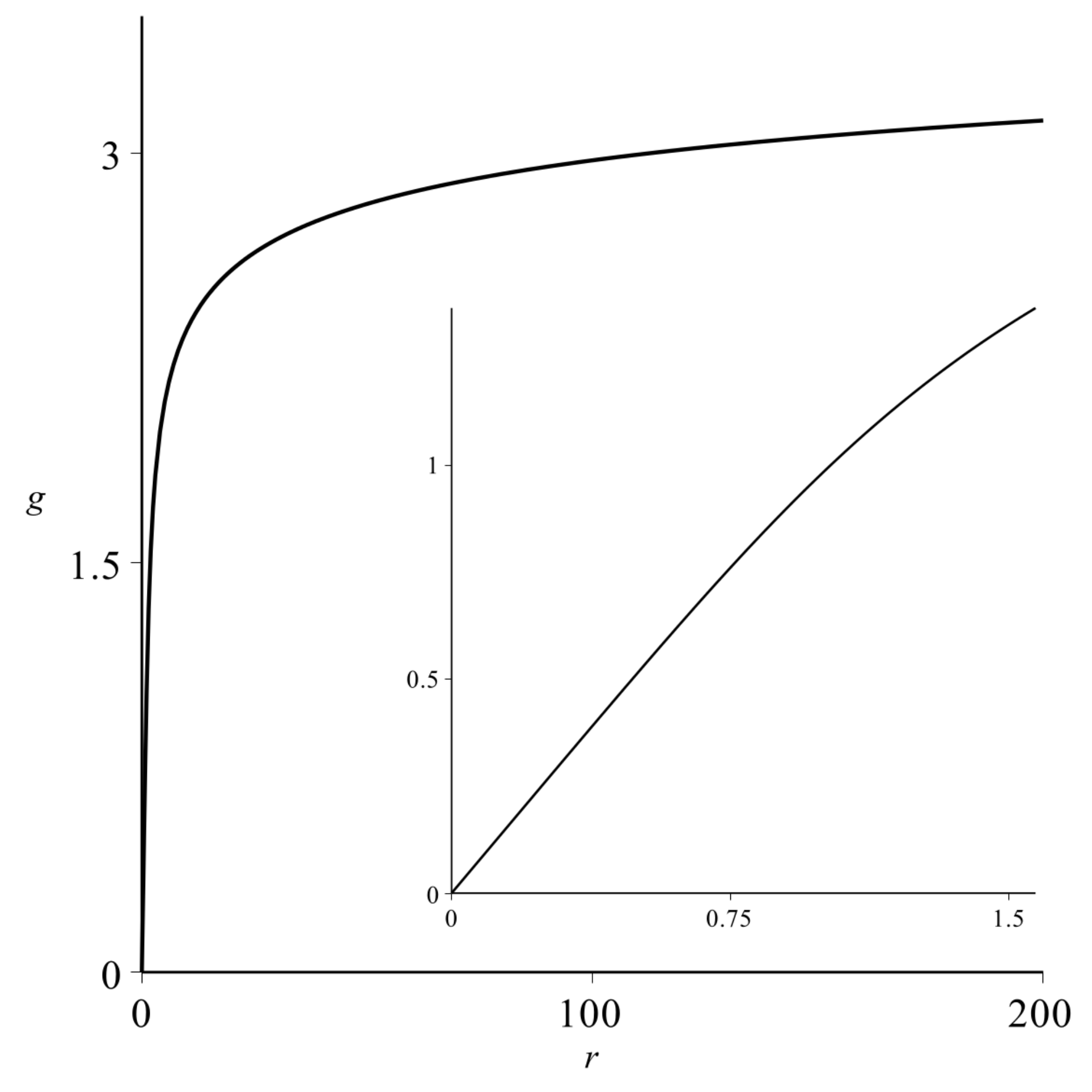}
\caption{The functions $a(r)$ (left) and $g(r)$ (right), solutions of Eqs.~\eqref{foex1cs}. The insets show the behavior near the origin, in the interval $r\in[0,1.57]$.}
\label{fig8}
\end{figure}

The first order equations \eqref{focs} in this case reads
\bal\label{foex1cs}
g^\prime &= \frac{ag}{r},\\
\frac{a^\prime}{r} &=-g^2 \sech^2\left(\frac{g^2}{2}\right)\left(1-\tanh\left(\frac{g^2}{2}\right)\right).
\eal
We have not been able to find analytical solutions for them. However, the behavior of the solutions near the origin may be studied by taking $a(r)=1-a_0(r)$ and $g(r)=g_0(r)$, similarly to the previous sections. By substituting them in the above equations, we get that
\be\label{oriex1cs}
a_0(r)\propto r^4 \quad\text{and}\quad g_0(r)\propto r.
\ee
This helps as a guide in the numerical calculations. In Fig.~\ref{fig8}, we plot the solutions. In fact, we see the behavior of the functions near the origin as given above. These solutions behaves similarly to the ones in Maxwell dynamics: $g(r)$ goes to infinity as $r$ increases but $a(r)$ tends to zero very slowly, presenting a tail that goes far away from the origin. This feature is the opposite of the one found for compact Chern-Simons vortices in Ref.~\cite{compcs}.

\begin{figure}[htb!]
\centering
\includegraphics[width=4.2cm]{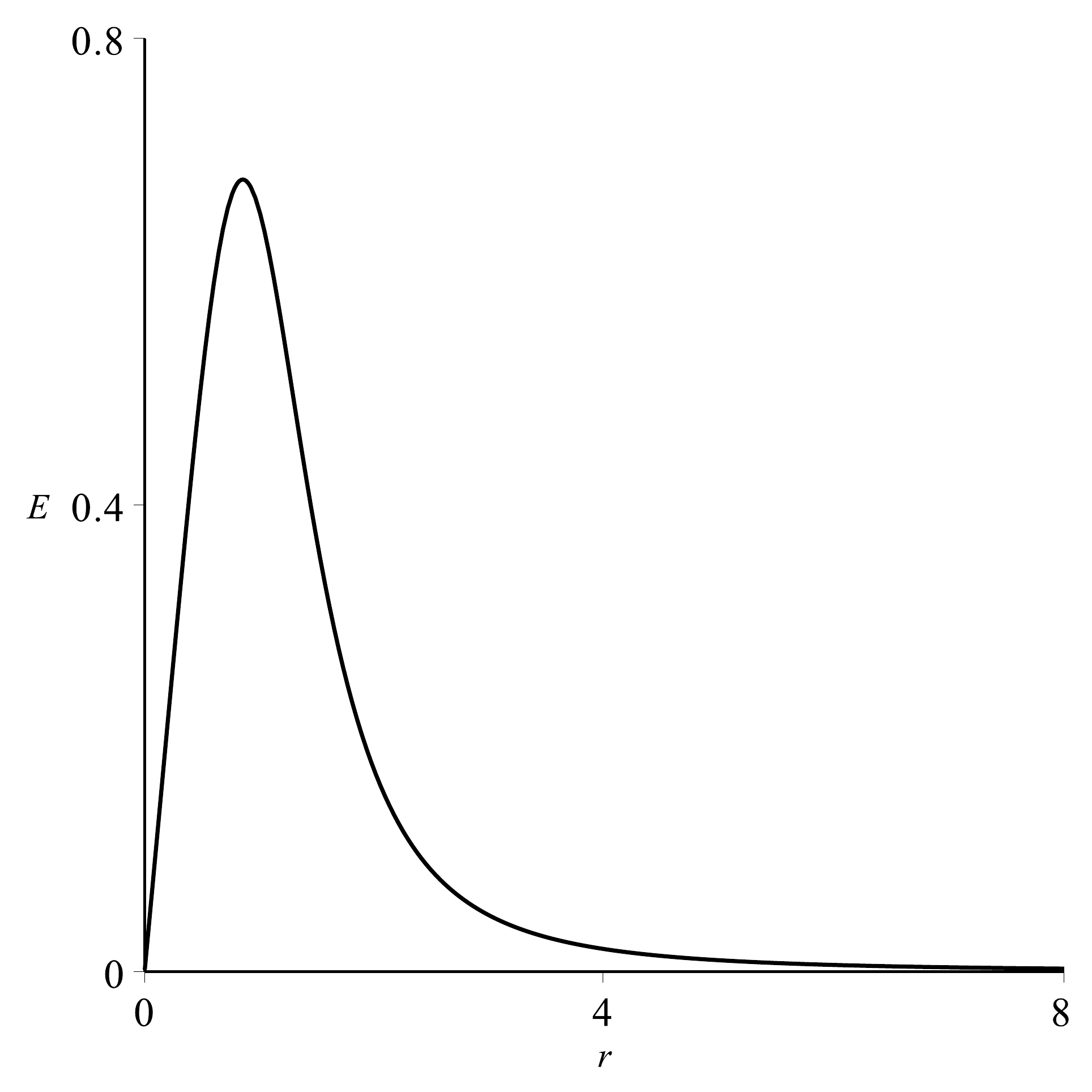}
\includegraphics[width=4.2cm]{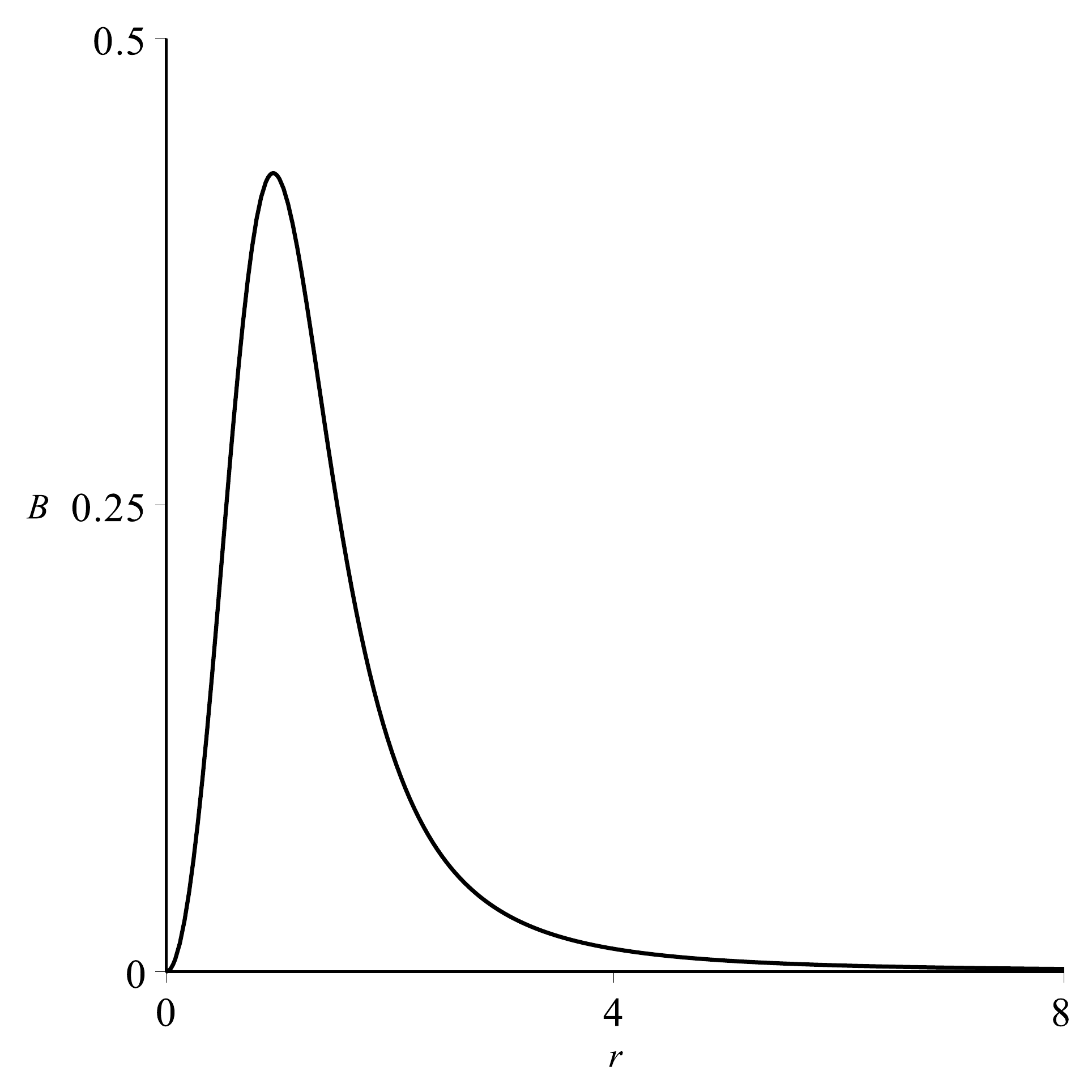}
\includegraphics[width=4.2cm]{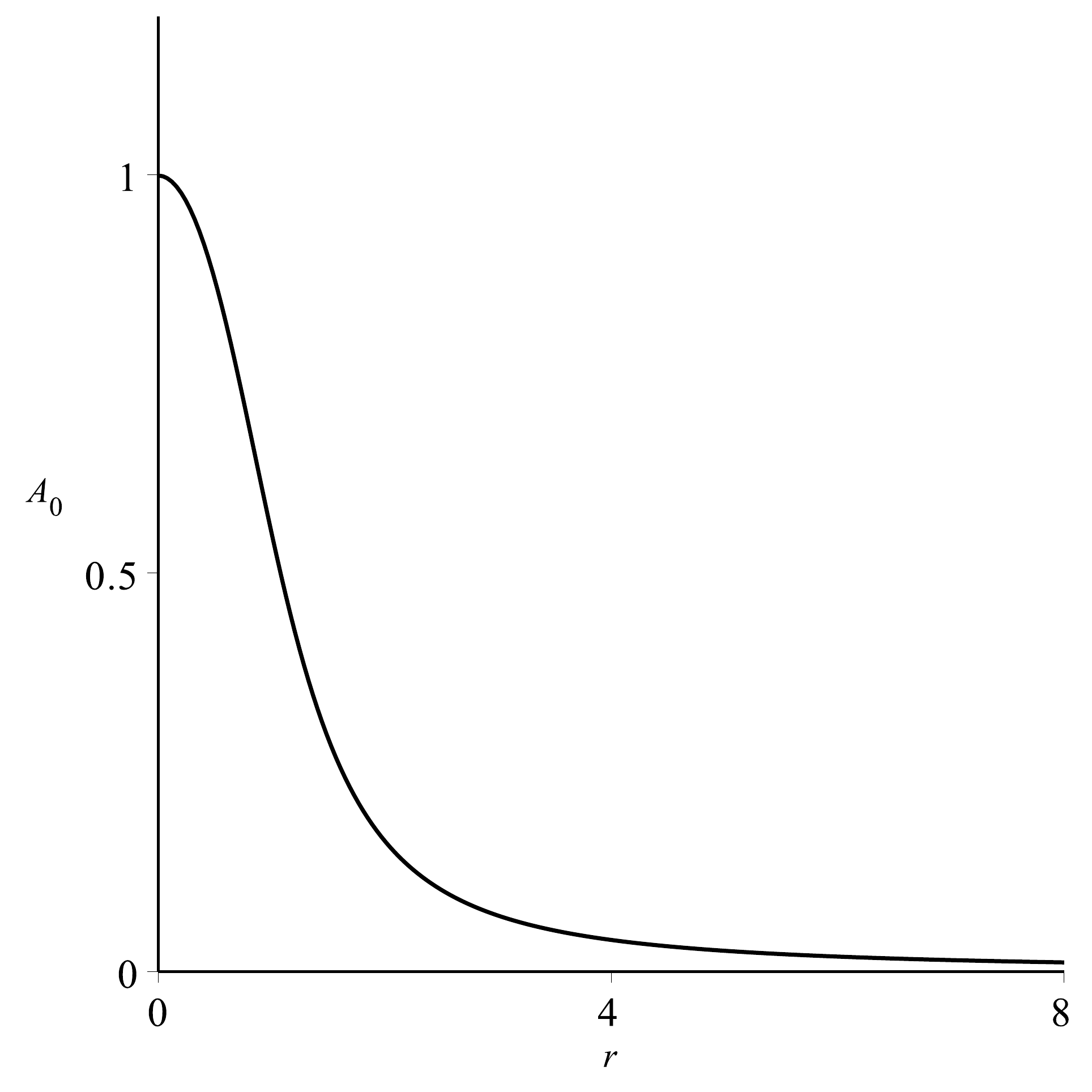}
\includegraphics[width=4.2cm]{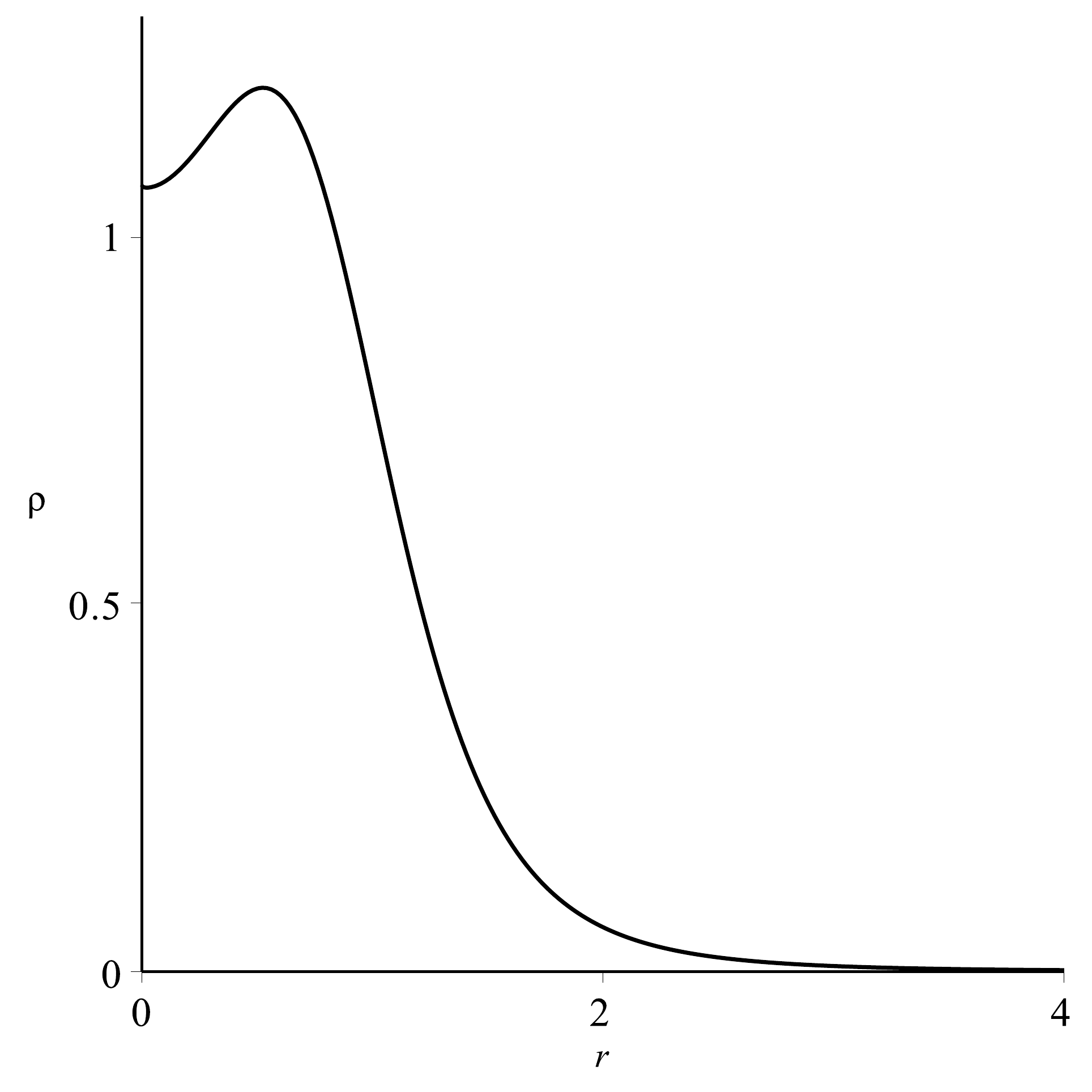}
\caption{The electric field (upper left), the magnetic field (upper right), the temporal gauge field component (bottom left) and the energy density (bottom right) for the solutions of Eqs.~\eqref{foex1cs}.}
\label{fig9}
\end{figure}

We now turn our attention to the auxiliar function $W(a,g)$ from Eq.~\eqref{wcs}. It is given by
\be\label{wex1cs}
W(a,g) = -a+a\tanh\left(\frac{g^2}{2}\right).
\ee
This is exactly the same function that appears in Eq.~\eqref{wex1m}. By using Eq.~\eqref{ewcs}, we get that the energy of the stressless solutions is $E=2\pi$. To calculate the electric field intensity and the magnetic field, one has to use the numerical solutions of Eqs.~\eqref{foex1cs} in Eqs.~\eqref{eb2cs}. The energy density must be calculated in a similar manner, by using the expression given below, which comes from Eq.~\eqref{rhocs}:
\be
\begin{aligned}
	\rho &=  \frac{{a^\prime}^2}{2r^2 g^2}\cosh^2\left(\frac{g^2}{2}\right) + \frac12\left({g^\prime}^2+\frac{a^2g^2}{r^2}\right)\,\sech^2\left(\frac{g^2}{2}\right) \\
	    &\hspace{4mm} + \frac12g^2\sech^2\left(\frac{g^2}{2} \right) \left(1-\tanh\left(\frac{g^2}{2}\right)\right)^2
\end{aligned}
\ee
In Fig.~\ref{fig9}, we plot the electric field, the magnetic field, the temporal component of the gauge field from Eq.~\eqref{A0} and the energy density.
As in the previous models, a numerical integration of the magnetic field and energy density gives the flux $\Phi\approx 2\pi$ and energy $\rho\approx2\pi$. Thus, the tail of the solutions does not seem to contribute to change the topological charge, since it is given by the flux. Therefore, in the Chern-Simons scenario, vortices in vacuumless systems have the topological current \eqref{topcurr} well defined that does not require any special definitions as done in Ref.~\cite{vbazeia} for kinks. 

\subsection{Second Model}
\begin{figure}[htb!]
\centering
\includegraphics[width=4.2cm]{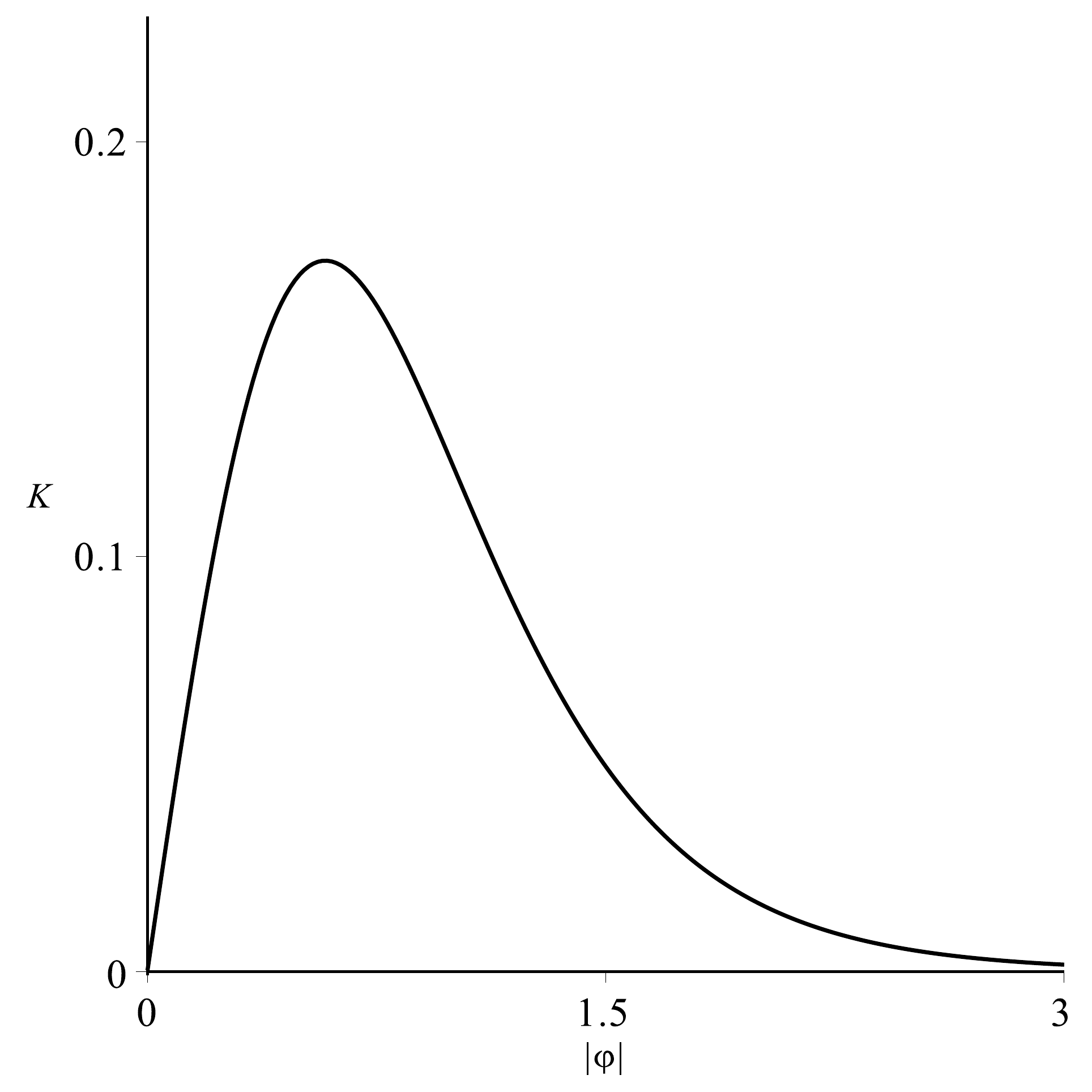}
\includegraphics[width=4.2cm]{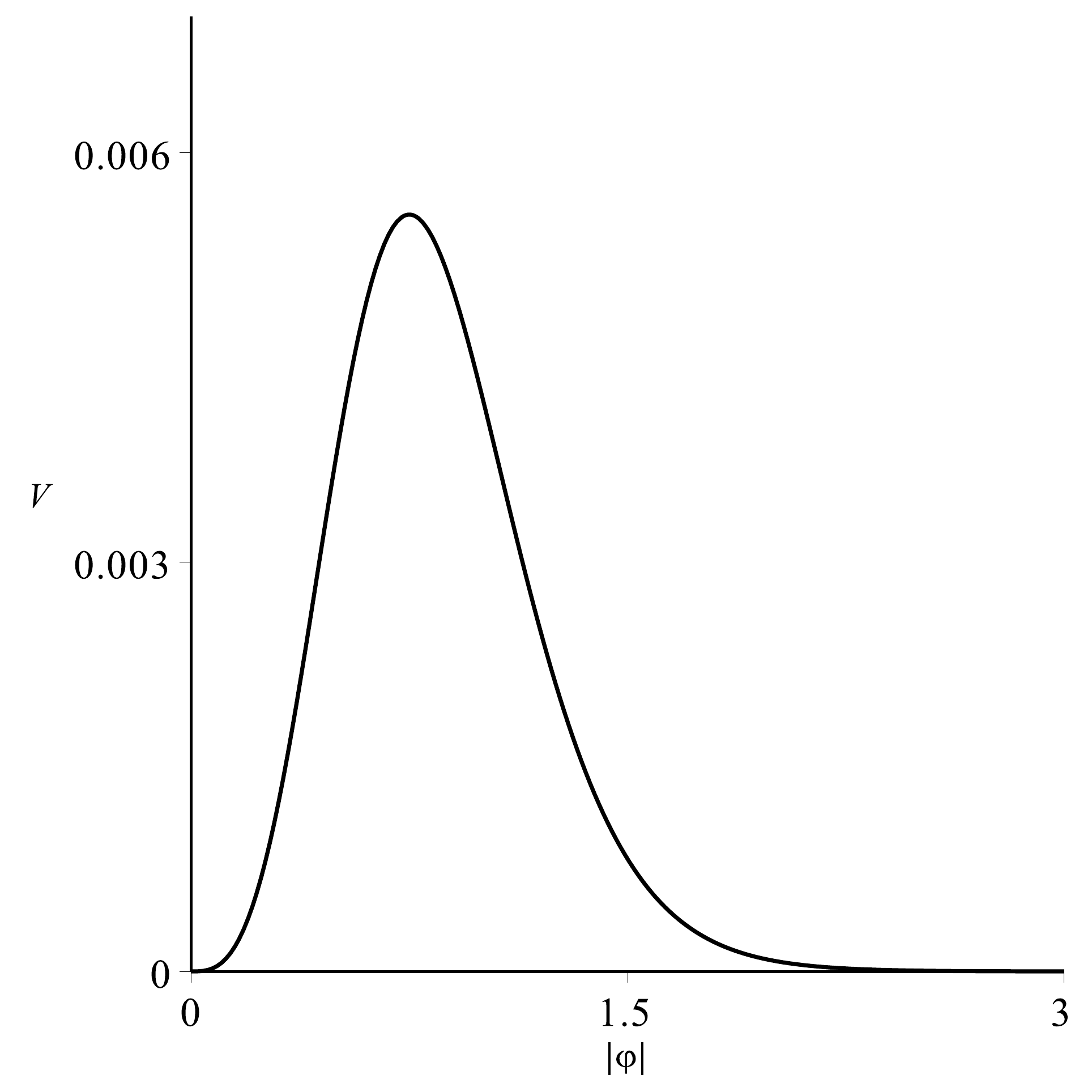}
\caption{The function $K(|\vphi|)$ (left) and the potential $V(|\vphi|)$ (right) given by Eqs.~\eqref{vk2cs}.}
\label{fig10}
\end{figure}

We now present a new model, given by the functions
\bes\label{vk2cs}
\bal
K(|\vphi|) &= \frac12 \frac{\sech^2(|\vphi|)\tanh^2(|\vphi|)}{|\vphi|}, \\
V(|\vphi|) &= \frac{1}{18} |\vphi| \,\sech^2(|\vphi|)\tanh^2(|\vphi|)\left( 1- \tanh^3(|\vphi|) \right)^2.
\eal
\ees
Differently of the previous model, the minima of both $K(|\vphi|)$ and the potential are located at $|\vphi|=0$ and $|\vphi|\to\infty$. The potential presents a maximum at $|\vphi_m| \approx 0.7500$, such that $V(|\vphi_m|) \approx 0.0055$. These features can be seen in Fig.~\ref{fig10}, in which we have plotted $K(|\vphi|)$ and the potential.

\begin{figure}[htb!]
\centering
\includegraphics[width=4.2cm]{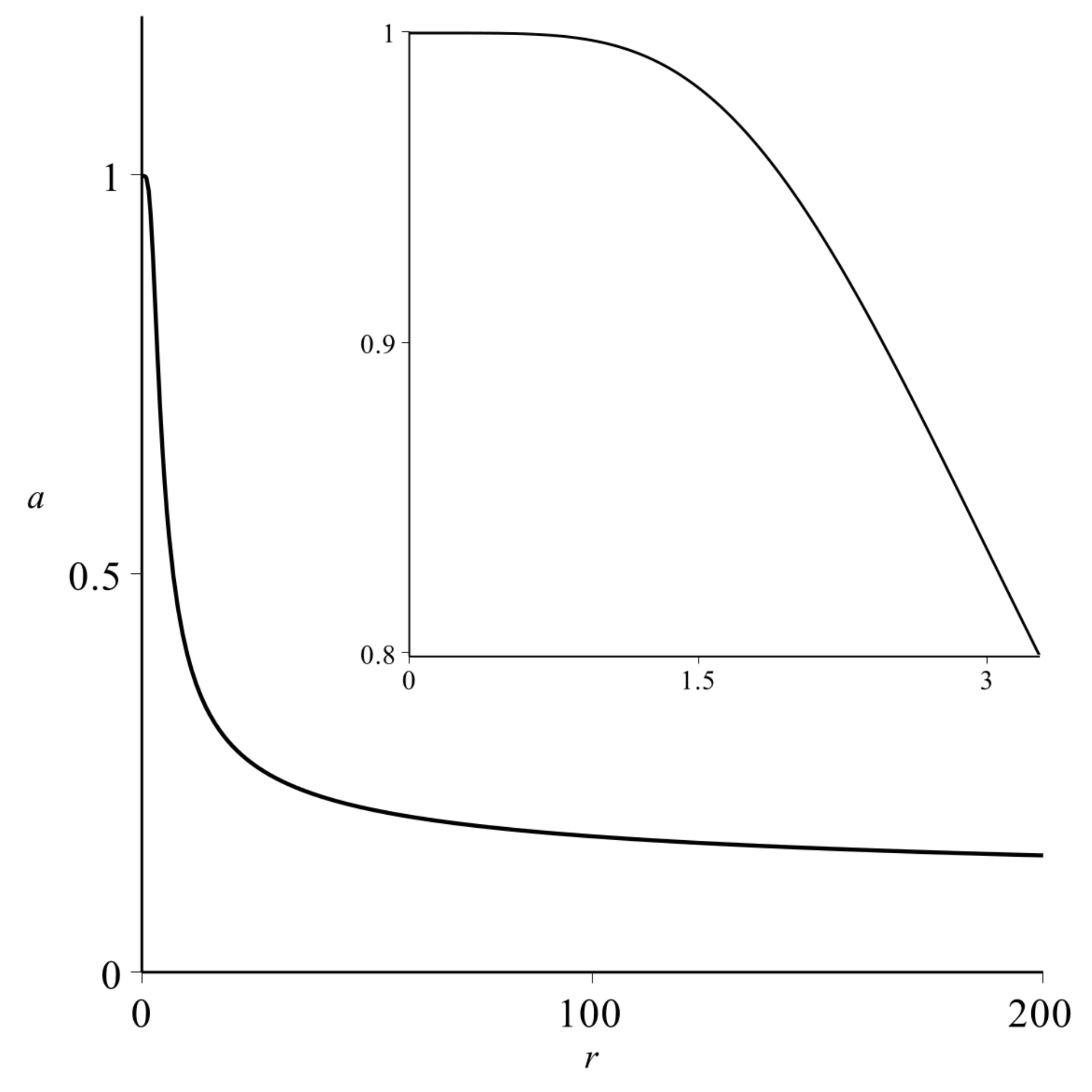}
\includegraphics[width=4.2cm]{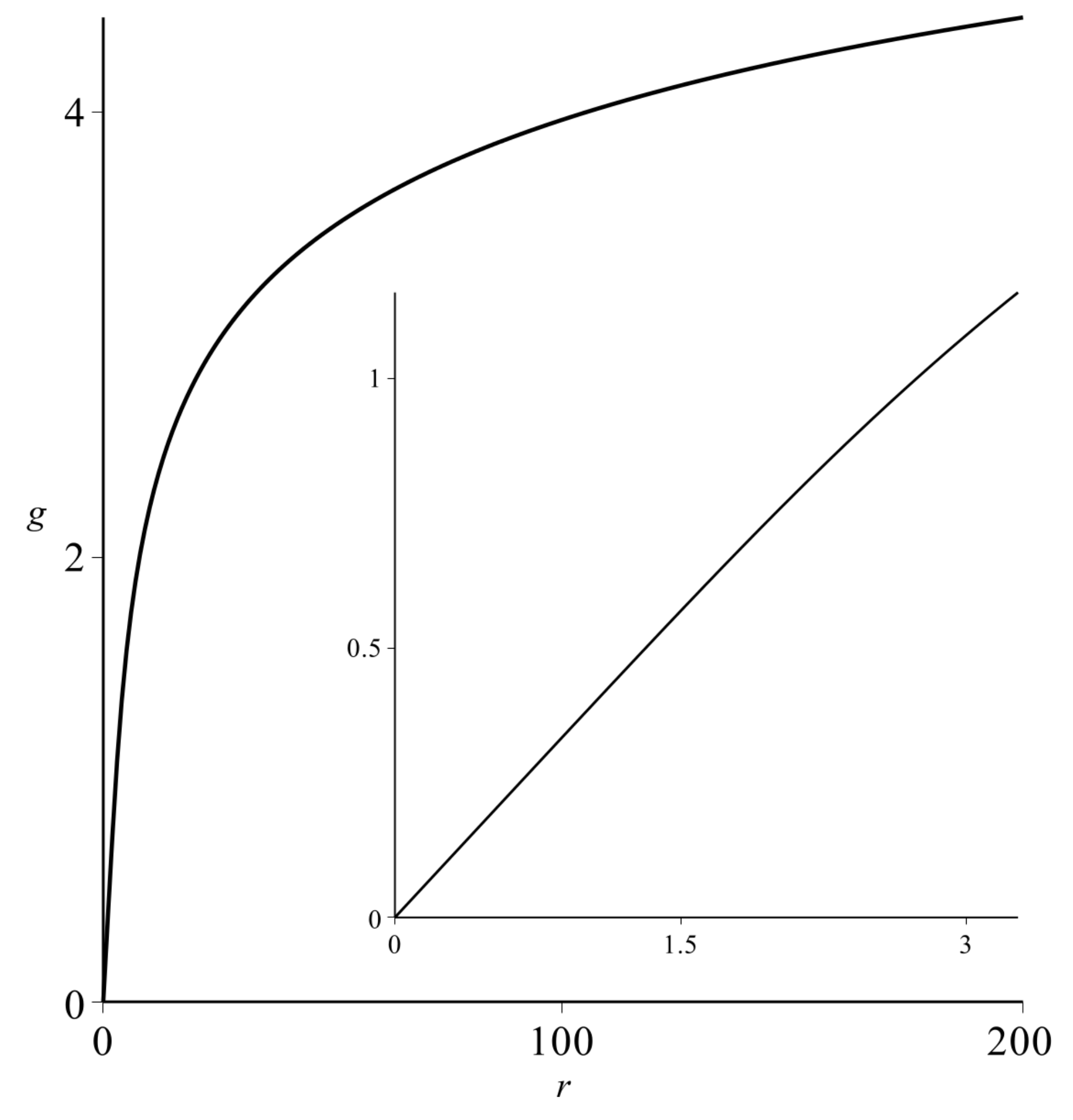}
\caption{The functions $a(r)$ (left) and $g(r)$ (right), solutions of Eqs.~\eqref{foex2cs}. The insets show the behavior near the origin, in the interval $r\in[0,3.27]$.}
\label{fig11}
\end{figure}

To calculate our solutions, we consider the first order equations \eqref{focs} to get
\bal\label{foex2cs}
g^\prime &= \frac{ag}{r},\\
\frac{a^\prime}{r} &=-\frac13 g\, \sech^2(g)\tanh^2(g)\left(1-\tanh^3(g)\right).
\eal
We have not been able to find the analytical solutions of the above equations. Nevertheless, it is worth to estimate their behavior near the origin by taking $a(r)=1-a_0(r)$ and $g(r)=g_0(r)$, similarly to was done before for the latter models. This approach leads to
\be\label{oriex2cs}
a_0(r)\propto r^5 \quad\text{and}\quad g_0(r)\propto r.
\ee
In Fig.~\ref{fig11} we plot the solutions of Eq.~\eqref{foex2cs}.
Notice that $a(r)$ is almost constant near the origin. This is due to the form of Eqs.~\eqref{oriex2cs}. As in the previous models, $g(r)$ tends to infinity as $r$ becomes larger and larger. Also, we see $a(r)$ tends to vanish very slow when $r\to\infty$, also presenting a tail which extends far away from the origin.

\begin{figure}[htb!]
\centering
\includegraphics[width=4.2cm]{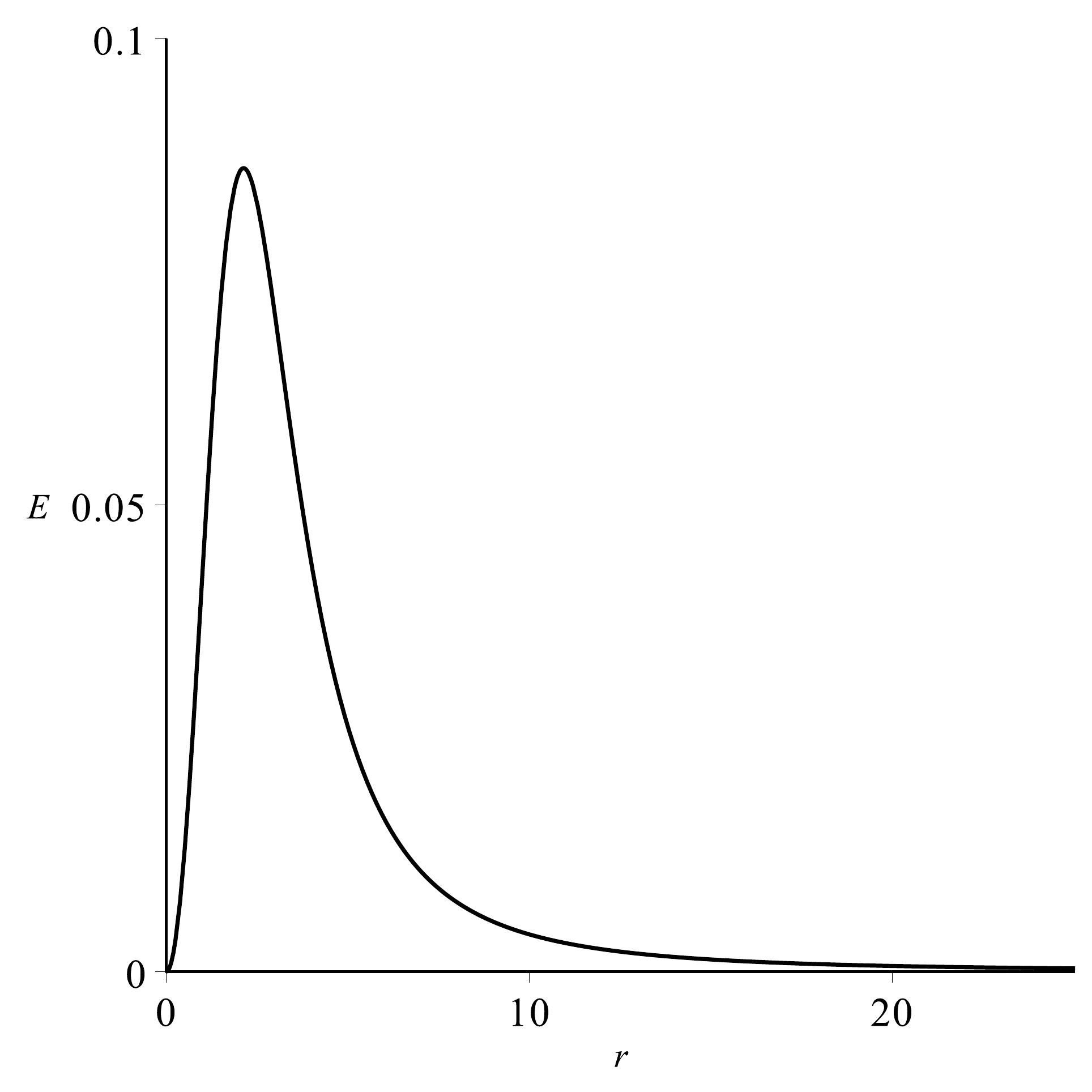}
\includegraphics[width=4.2cm]{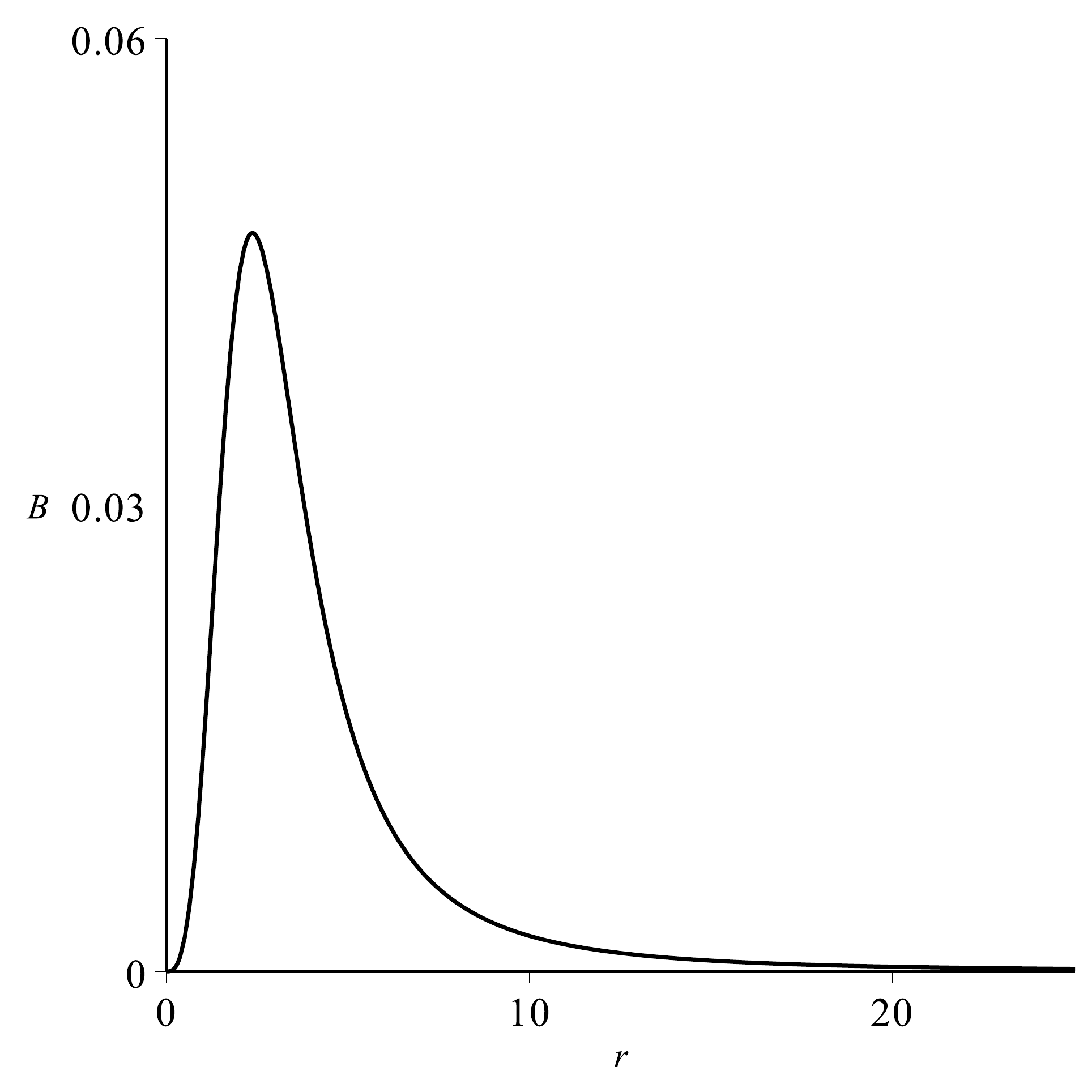}
\includegraphics[width=4.2cm]{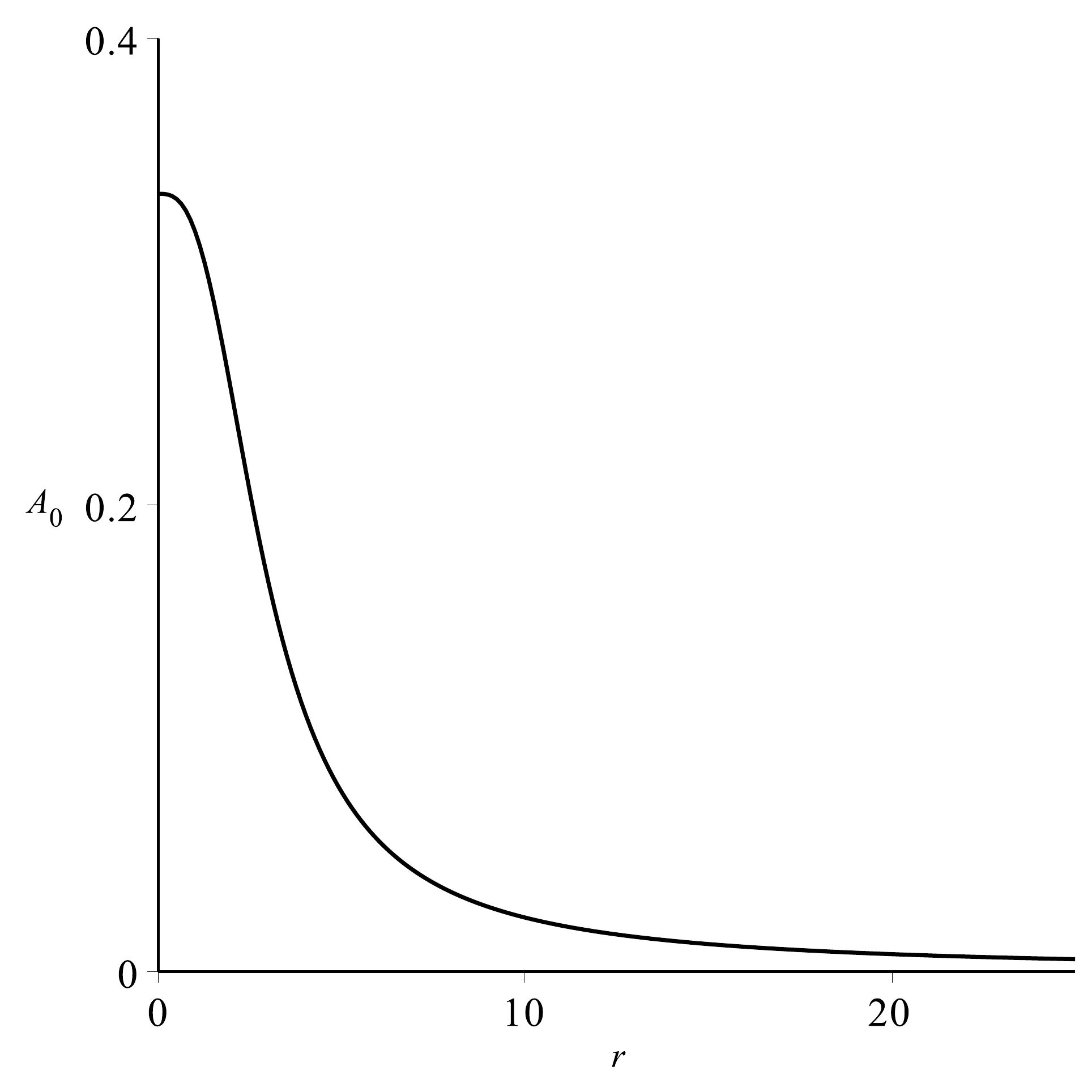}
\includegraphics[width=4.2cm]{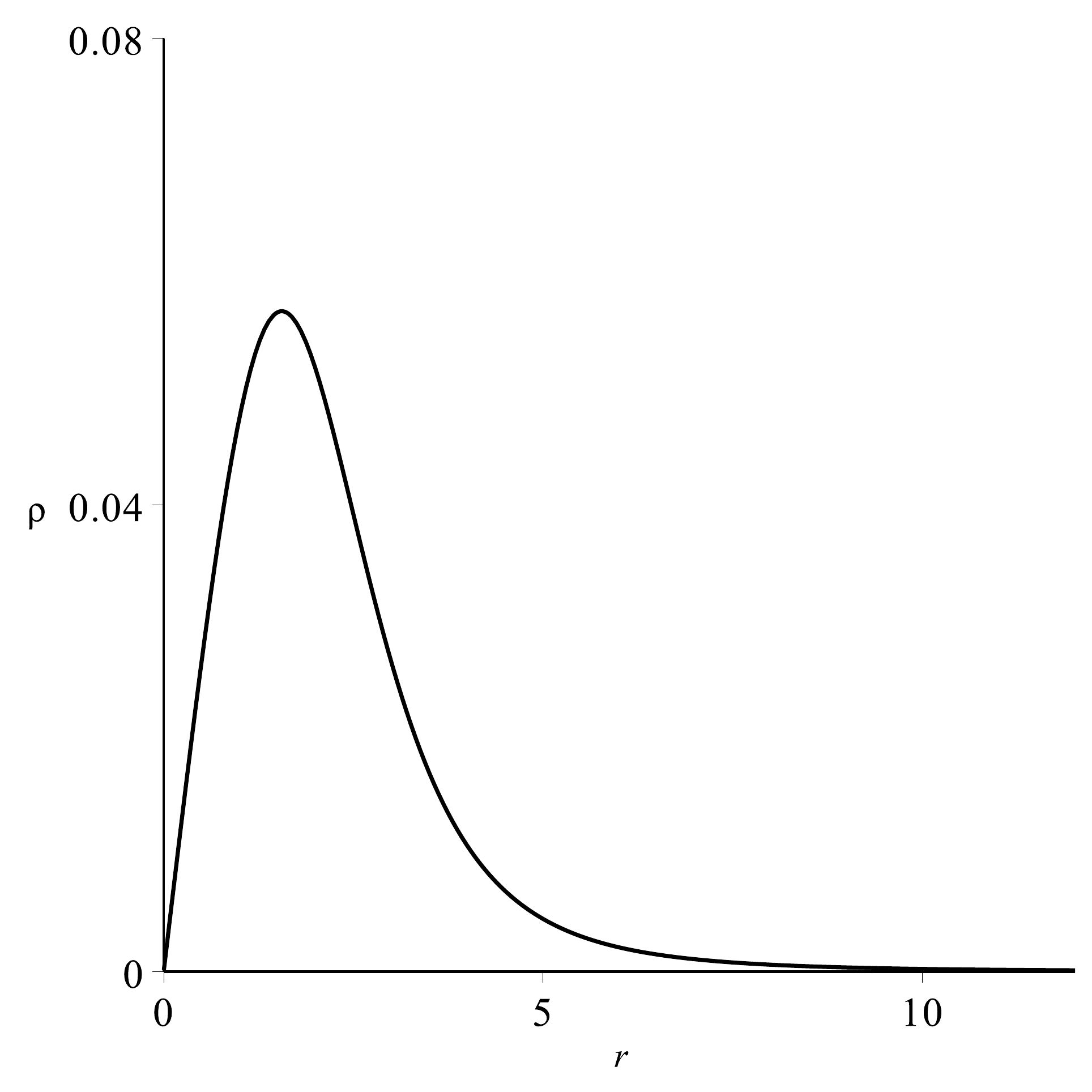}
\caption{The electric field (upper left), the magnetic field (upper right), the temporal gauge component (bottom left) and the energy density (bottom right) for the solutions of Eqs.~\eqref{foex2cs}.}
\label{fig12}
\end{figure}

In this case, the function $W(a,g)$ in Eq.~\eqref{wcs} becomes
\be\label{wex2cs}
W(a,g) = \frac{a}{3} \left(1-\tanh^3(g)\right).
\ee
Therefore, by using Eq.~\eqref{ewcs}, we conclude that the energy is $E=2\pi/3$. To calculate the intensity of the electric and magnetic fields, one has to use the numerical solutions into Eq.~\eqref{eb2cs}. The same occurs to evaluate the energy density, which comes from Eq.~\eqref{rhocs} that leads to
\be
\begin{aligned}
	\rho &=  \frac{{a^\prime}^2}{2r^2 g}\cosh^2(g)\coth^2(g) \\
	     &\hspace{4mm}+  \left({g^\prime}^2+\frac{a^2g^2}{r^2}\right) \frac{\sech^2(g)\tanh^2(g)}{2g}\\
	     &\hspace{4mm}+\frac{1}{18} g \,\sech^2(g)\tanh^2(g)\left( 1- \tanh^3(g) \right)^2.
\end{aligned}
\ee
In Fig.~\ref{fig12}, we plot the electric and magnetic fields, the temporal gauge component \eqref{A0} and the above energy density.
As for all of our previous models, the topological charge, given by the flux, remains unchanged from Eq.~\eqref{fluxm}, having the value $\Phi\approx 2\pi$ obtained from a numerical integration. The energy can be obtained numerically and it is given by $E\approx2\pi/3$, the same value obtained from the function $W(a,g)$ of Eq.~\eqref{wex2cs}. Also, we see the energy density in this model present a valley deeper than in the previous one. 

\section{Conclusions}\label{sec4}
In this work, we have investigated vortices in vacuumless systems with Maxwell and Chern-Simons dynamics. In both scenarios, we have studied the properties of the generalized models in the classes \eqref{lmax} and \eqref{lcs} and, following Ref.~\cite{godvortex}, we have used a first order formalism that allows to calculate the energy without knowing the explicit form of the solutions.

The behavior of the potentials are different at $|\vphi|=0$, depending on the scenario; in the Maxwell case, they are nonvanishing, whilst in the Chern-Simons models they are zero. The hole around the origin in the potentials for the Chern-Simons dynamics makes the magnetic field vanish at $r=0$. Regardless the differences in the behavior of the magnetic field, the magnetic flux is always quantized by the vorticity $n$. Furthermore, even though we have worked only with $n=1$, for simplicity, in our examples, it is worth commenting that we have checked the energy is also quantized by the vorticity, $n$.

An interesting result is that the vortex solutions in vacuumless systems present a large tail that extends far away from the origin. The scalar field is asymptotically divergent and has infinite amplitude. Then, the solutions looses the locality. However the electric field, if it exists, the magnetic field, as well as the energy density, are localized. This avoids the possibility of having infinite energies and fluxes. The flux is well defined and still works as a topological invariant. Unlike the kinks, we concluded that vortices in vacuumless systems do not require any special definition of the topological current to study its topological character. 

We then discovered vortices with a new behavior, whose solutions present a long tail. We hope these results encourage new research in the area, stimulating the study of new models in this and other contexts. One can follow the direction of Ref.~\cite{fermion} and study the demeanor of fermions in the background of these vortex structures. Also, the collective behavior of these vortices seems of interest, since it may give rise to non-standard interactions due to the particular aforementioned features of the solutions. Furthermore, following the lines of Ref.~\cite{vac3}, one also can study the gravitational field of these vortices. Other perspective is to investigate these structures in models with enlarged symmetries \cite{hidden1,hidden2,hidden3,shif1,shif2,shif3,vortexint}, which may make them appear in the hidden sector, for instance. Finally, one may try to extend the current investigation to other topological structures, such as monopoles \cite{mono1,mono2}, and non-topological structures, such as lumps \cite{lump1,lump2,lump3} and Q-balls \cite{qball1,qball2}. Some of these issues are under consideration and will be reported on the near future.


\acknowledgements{We would like to thank Dionisio Bazeia and Roberto Menezes for the discussions that have contributed to this work. We would also like to acknowledge the Brazilian agency CNPq, research project 155551/2018-3, for the financial support.}


\end{document}